\documentclass[aps,pra,twocolumn,showpacs]{revtex4-1}

\usepackage{graphicx}
\usepackage{dcolumn}
\usepackage{bm}
\usepackage[version=3]{mhchem}
\usepackage{color}

\begin{document}

\title{Defect Physics, Delithiation Mechanism, and Electronic and Ionic Conduction in Layered Lithium Manganese Oxide Cathode Materials}
\author{Khang Hoang}
\email[E-mail: ]{khang.hoang@ndsu.edu}
\affiliation{Center for Computationally Assisted Science and Technology, North Dakota State University, Fargo, ND 58108, USA.}

\date{\today}

\begin{abstract}

Layered \ce{LiMnO2} and \ce{Li2MnO3} are of great interest for lithium-ion battery cathodes because of their high theoretical capacities. The practical application of these materials is, however, limited due to poor electrochemical performance. We herein report a comprehensive first-principles study of defect physics in \ce{LiMnO2} and \ce{Li2MnO3} using hybrid-density functional calculations. We find that manganese antisites have low formation energies in \ce{LiMnO2} and may act as nucleation sites for the formation of impurity phases. The antisites can also occur with high concentrations in \ce{Li2MnO3}; however, unlike in \ce{LiMnO2}, they can be eliminated by tuning the experimental conditions during preparation. Other intrinsic point defects may also occur and have an impact on the materials' properties and functioning. An analysis of the formation of lithium vacancies indicates that lithium extraction from \ce{LiMnO2} is associated with oxidation at the manganese site, resulting in the formation of manganese small hole polarons; whereas in \ce{Li2MnO3} the intrinsic delithiation mechanism involves oxidation at the oxygen site, leading to the formation of bound oxygen hole polarons $\eta_{\rm O}^{+}$. The layered oxides are found to have no or negligible bandlike carriers and they cannot be doped n- or p-type. The electronic conduction proceeds through hopping of hole and/or electron polarons; the ionic conduction occurs through lithium monovacancy and/or divacancy migration mechanisms. Since $\eta_{\rm O}^{+}$ is not stable in the absence of negatively charged lithium vacancies in bulk \ce{Li2MnO3}, the electronic conduction near the start of delithiation is likely to be poor. We suggest that the electronic conduction associated with $\eta_{\rm O}^{+}$ and, hence, the electrochemical performance of \ce{Li2MnO3} can be improved through nanostructuring and/or ion substitution.

\end{abstract}


\maketitle


\section{Introduction}\label{sec;intro}

Layered lithium manganese oxides \ce{LiMnO2} and \ce{Li2MnO3} have been considered for lithium-ion battery cathodes as manganese is inexpensive and environmentally benign compared to cobalt and nickel and the materials have high theoretical capacities \cite{Ellis2010,He2012}. \ce{Li2MnO3} is also an important component in high-capacity $x$\ce{Li2MnO3}$\cdot$(1$-$$x$)\ce{LiMO2} (M = Mn, Ni, Co, etc.) cathode materials \cite{Thackeray2007}. Layered \ce{LiMnO2}, in which cation planes alternate as lithium layers and manganese layers, is prepared by an ion-exchange method \cite{Armstrong1996,Capitaine1996} as the synthesis at high temperatures often results in an orthorhombic phase. The compound exhibits strong cation mixing \cite{Armstrong1996} and poor electrochemical performance which is usually ascribed to the transformation into spinel \ce{LiMn2O4} during charge-discharge cycling \cite{Vitins1997}. Li-excess \ce{Li2MnO3}, also known as Li[Li$_{1/3}$Mn$_{2/3}$]O$_{2}$, has a layered structure similar to that of \ce{LiMnO2} but one-third of the ions in the manganese layer are replaced by lithium ions \cite{Massarotti1997}. Since in \ce{Li2MnO3} manganese exists as Mn$^{4+}$, the material was initially believed to be electrochemically inactive. It was later demonstrated that \ce{Li2MnO3} can be made electrochemically active by acid leaching \cite{Rossouw1991} or charging to high voltages \cite{Kalyani1999}. The material shows very limited electrochemical capacity, which could be due to poor kinetics of lithium extraction and reinsertion \cite{Sathiya2013}. However, it is also reported that the electrochemical performance of \ce{Li2MnO3} can be improved through nanostructuring \cite{Jain2005,Yu2009,Lim2012,Wu2014} or ion substitution \cite{Sathiya2013}. 

Several different mechanisms have been proposed to explain the unconventional lithium-extraction behavior of layered \ce{Li2MnO3}. These involve either (i) the oxidation of Mn$^{4+}$ to Mn$^{5+}$ \cite{Kalyani1999}, (ii) simultaneous removal of lithium and oxygen \cite{Lu2002,Armstrong2005}, (iii) oxidation of Mn$^{3+}$ associated with oxygen deficiency to Mn$^{4+}$ \cite{Pasero2005,Okamoto2011,Kubota2012}, (iv) oxidation of the electrolyte and exchange of H$^{+}$ for Li$^{+}$ \cite{Robertson2002,Armstrong2005,Benedek2008}, or (v) oxygen oxidation \cite{Kim2004,Koga2013,Sathiya2013,Koyama2009}. Regarding the final mechanism, direct experimental evidence for the reversibility of the O$^{2-}$ to O$^-$ anionic process upon cycling has been observed in x-ray photoemission spectroscopy studies of \ce{Li2Ru$_{1-y}$Mn$_y$O3} ($0.2 \leq y \leq 0.8$) cathode materials \cite{Sathiya2013}. On the theory side, although oxygen oxidation in Li$_{2-x}$MnO$_{3}$ is mentioned in several previous computational works \cite{Koyama2009,Xiao2012,Lee2014}, the formation of O$^-$ has never been clearly demonstrated, especially at small $x$ values. Apparently, further theoretical and computational studies are needed in order to fully understand the properties of \ce{Li2MnO3} and related materials. As demonstrated in our previous works \cite{Hoang2011,Hoang2014,HoangLiMn2O4}, first-principles defect calculations based on density-functional theory (DFT) can serve as an important tool in this regard. 

We herein present a comprehensive computational approach based on state-of-the-art first-principles defect calculations to studying battery-electrode materials. In this approach, we start with an investigation of the bulk properties of and phase diagrams associated with the host compounds and then proceed with a detailed investigation of the structure, energetics, and migration of all possible intrinsic electronic and ionic point defects in the materials. An expression for the lithium-extraction voltage is also derived based on an expression for the formation energy of lithium vacancies. We then illustrate how this approach helps uncover the defect physics, intrinsic mechanisms for the delithiation (and lithiation), and electronic and ionic conduction mechanisms in layered oxide materials \ce{LiMnO2} and \ce{Li2MnO3}. Most interestingly, we find that in \ce{Li2MnO3} the lithium-extraction process is associated with oxidation at the oxygen site, instead of the transition-metal site as in other complex oxide electrode materials, leading to the formation of bound oxygen hole polarons. In light of our results, we provide explanations for the experimental observations, guidelines for defect-controlled synthesis and defect characterization, suggestions for improving the electronic conduction, and ultimately insights for designing high-capacity battery-electrode materials.

\section{Methodology}\label{sec;method}

\subsection{Hybrid functional calculations}

Our calculations for the bulk properties and point defects are based on DFT, using the Heyd-Scuseria-Ernzerhof (HSE06) screened hybrid functional \cite{heyd:8207,paier:154709}, the projector augmented wave method \cite{PAW1,PAW2}, and a plane-wave basis set, as implemented in the Vienna {\it Ab Initio} Simulation Package (VASP) \cite{VASP1,VASP2,VASP3}. In these calculations, we set the Hartree-Fock mixing parameter and the screening length to the standard values of 0.25 and 10 {\AA}, respectively. The use of the HSE06 hybrid functional, where all orbitals are treated on equal footing, is to ensure the transferability of calculations across compounds in the Li-Mn-O phase diagram and that the physics of the complex transition-metal oxides is properly described \cite{Hoang2014,HoangLiMn2O4}. We note that the HSE06 functional has also been employed in the study of polarons in other battery-electrode materials \cite{Ong2011,Johannes2012,Ong2012}. The GGA+$U$ method \cite{dudarev1998,liechtenstein1995}, an extension of the generalized-gradient approximation (GGA) within DFT \cite{GGA}, is used only for comparison in some specific calculations. In these GGA+$U$ calculations, the on-site Hubbard corrections are applied to both the Mn 3d states and O 2p states. 

Intrinsic point defects in \ce{LiMnO2} and \ce{Li2MnO3} are treated within the supercell approach, in which a defect is included in a finite volume of the host material and this structure is periodically repeated. For the defect calculations, we use hexagonal supercells containing 108 atoms per cell, and integrations over the Brillouin zone are carried out using the $\Gamma$ point. The plane-wave basis-set cutoff is set to 500 eV. Convergence with respect to self-consistent iterations is assumed when the total-energy difference between cycles is less than 10$^{-4}$ eV and the residual forces are less than 0.01 eV/{\AA}. In these defect calculations, which are performed with spin polarization and the ferromagnetic spin configuration for the manganese array in the lattice, the lattice parameters are fixed to the calculated bulk values but all the internal coordinates are fully relaxed.

\subsection{Defect-formation energies}\label{sec;formenergy}

The properties of a point defect in solids are characterized by its formation energy and migration barrier. In our calculations, the latter is calculated by using the climbing-image nudged elastic-band (NEB) method \cite{ci-neb}; the former is computed using the total energies from DFT calculations. The formation energy of a defect X in charge state $q$ is defined as \cite{walle:3851}
\begin{equation}\label{eq:eform}
E^f({\mathrm{X}}^q)=E_{\mathrm{tot}}({\mathrm{X}}^q)-E_{\mathrm{tot}}({\mathrm{bulk}})-\sum_{i}{n_i\mu_i}+q(E_{\mathrm{v}}+\mu_{e})+ \Delta^q ,
\end{equation}
where $E_{\mathrm{tot}}(\mathrm{X}^{q})$ and $E_{\mathrm{tot}}(\mathrm{bulk})$ are, respectively, the total energies of a supercell containing the defect X and of an equivalent supercell of the perfect bulk material. The integer $n_{i}$ indicates the number of atoms of species $i$ that have been added to ($n_{i}$$>$0) or removed from ($n_{i}$$<$0) the supercell to form the defect; $\mu_{i}$ is the atomic chemical potential of species $i$, representing the energy of the reservoir with which atoms are being exchanged, and is referenced to the bulk metals or O$_{2}$ molecules at 0 K. $\mu_{e}$ is the electronic chemical potential or the Fermi level, representing the energy of the electron reservoir, referenced to the valence-band maximum in the bulk ($E_{\mathrm{v}}$). $\Delta^q$ is the correction term to align the electrostatic potentials of the bulk and defect supercells and to account for finite-supercell-size effects on the total energies of charged defects \cite{walle:3851}. In this work, we adopt the approach of Freysoldt {\it et al.}~\cite{Freysoldt,Freysoldt11}, in which the correction term $\Delta^q$ to the formation energies of charged defects is determined without empirical parameters. 

The concentration of a defect at temperature $T$ is related to its formation energy through the expression \cite{walle:3851} 
\begin{equation}\label{eq:concen} 
c=N_{\mathrm{sites}}N_{\mathrm{config}}\mathrm{exp}\left(\frac{-E^{f}}{k_{B}T}\right), 
\end{equation} 
where $N_{\mathrm{sites}}$ is the number of high-symmetry sites in the lattice per unit volume on which the defect can be incorporated, $N_{\mathrm{config}}$ is the number of equivalent configurations (per site), and $k_{B}$ is Boltzmann's constant. The energy in Eq.~(\ref{eq:concen}) is, in principle, a free energy; however, the entropy and volume terms are often neglected because they are negligible at relevant experimental conditions \cite{walle:3851}. This expression is valid in the dilute defect limit, i.e., neglecting the defect-defect interaction, and in thermodynamic equilibrium. As discussed in Ref.~\cite{walle:3851}, Eq.~(\ref{eq:concen}) is also applicable under conditions that are close to equilibrium or when the relevant defects are mobile enough to allow for equilibration at the temperatures of interest. It emerges from this expression that defects with low formation energies will easily form and occur in high concentrations. Furthermore, defect-formation energies should be positive; otherwise, the host compound would be unstable.

The atomic chemical potentials $\mu_{i}$ in Eq.~(\ref{eq:eform}) are variables and subject to thermodynamic constraints. The stability of the \ce{LiMnO2} phase, for example, requires
\begin{equation}\label{eq;limno2} 
\mu_{\rm Li}+\mu_{\rm Mn}+2\mu_{\rm O}=\Delta H^{f}({\rm LiMn}{\rm O}_{2}), 
\end{equation}
where $\Delta H^{f}$ is the formation enthalpy. Similarly, it is required that the atomic chemical potentials in the case of \ce{Li2MnO3} satisfy the condition
\begin{equation}\label{eq;li2mno3} 
2\mu_{\rm Li}+\mu_{\rm Mn}+3\mu_{\rm O}=\Delta H^{f}({\rm Li}_{2}{\rm Mn}{\rm O}_{3}).
\end{equation}
These conditions place a lower bound on the value of $\mu_{i}$. In addition, one needs to avoid precipitating bulk Li and Mn phases, or forming O$_{2}$ gas. These constraints set an upper bound on the atomic chemical potentials: $\mu_{i}$$\leq$0. There are, however, further thermodynamic constraints imposed by other competing Li-Mn-O phases which often place stronger bounds on $\mu_{i}$. For example, in order to avoid the formation of \ce{Li2O}, 
\begin{equation}\label{eq;li2o} 
2\mu_{\rm Li}+\mu_{{\rm O}}\leq \Delta H^{f}({\rm Li}_{2}{\rm O}). 
\end{equation}
By taking into account the constraints imposed by all competing phases, one can determine the range of Li, Mn, and O chemical-potential values in which the host compound \ce{LiMnO2} or \ce{Li2MnO3} is thermodynamically stable.

The oxygen chemical potential $\mu_{\rm O}$ can also be related to temperatures and pressures via the expression \cite{Reuter2001}: 
\begin{equation}\label{eq;muO} 
\mu_{\mathrm{O}}(T,p)=\mu_{\mathrm{O}}(T,p_{\circ}) + \frac{1}{2}k_{B}T {\rm ln}\frac{p}{p_{\circ}}, 
\end{equation} 
where $p$ and $p_{\circ}$ are, respectively, the partial pressure and reference partial pressure of \ce{O2} gas. This expression allows the calculation of $\mu_{\mathrm{O}}(T,p)$ if one knows the temperature dependence of $\mu_{\mathrm{O}}(T,p_{\circ})$ at a particular pressure $p_{\circ}$. In this work, the reference state of $\mu_{\mathrm{O}}(T,p)$ is chosen to be half of the total energy of an isolated \ce{O2} molecule at 0 K. In DFT calculations using the HSE06 functional, the binding energy of \ce{O2} with respect to spin-polarized O atoms is found to be 5.16 eV \cite{Hoang2014}, in good agreement with the experimental value of 5.12 eV \cite{chase}.

It should be noted that the Fermi level $\mu_{e}$ is not a free parameter. In principle, Eqs.~(\ref{eq:eform}) and (\ref{eq:concen}) can be written for every intrinsic defect and impurity in the material. The complete problem, including free-carrier concentrations in valence and conduction bands, if present, can then be solved self-consistently by imposing the charge-neutrality condition \cite{walle:3851}:
\begin{equation}\label{eq:neutrality}
\sum_{i}c_{i}q_{i}-n_{e}+n_{h}=0,
\end{equation}
where $c_{i}$ and $q_{i}$ are the concentration and charge, respectively, of defect or impurity X$_{i}$; $n_{e}$ and $n_{h}$ are free electron and hole concentrations, respectively; and the summation is over all defects and impurities.

\subsection{Lithium-extraction voltage}\label{sec;voltage}

In lithium-ion batteries, lithium ions Li$^{+}$ are extracted from the positive electrodes. For example, in \ce{LiMnO2} cathodes, the delithiation reaction occurs as
\begin{equation}\label{eq:oxidation}
{\rm LiMnO}_{2} \rightarrow {\rm Li}_{1-x}{\rm MnO}_{2} + x{\rm Li}^{+} + xe^{-}.  
\end{equation}
The liberated Li$^{+}$ ions then dissolve into the electrolyte. Lithium extraction can therefore be described in terms of the creation of lithium vacancies. In fact, as discussed later in Sec.~\ref{sec;formation}, the removal of a Li atom, i.e., Li$^{+} + e^{-}$, is equivalent to the formation of a lithium vacancy, hereafter denoted as $V_{\rm Li}^{0}$, in the electrode material.

According to Eq.~(\ref{eq:eform}), the formation energy of $x$ thermally activated lithium vacancies is given by
\begin{equation}\label{eq:nvli}
E^f(xV_{\mathrm{Li}}^0)=E_{\mathrm{tot}}(xV_{\mathrm{Li}}^0)-E_{\mathrm{tot}}({\mathrm{bulk}})+x\mu_{\rm Li}^{\ast},
\end{equation}
where $E_{\rm tot}$($xV_{\rm Li}^{0}$) is the total energy of the supercell containing the vacancies, e.g., the ${\rm Li}_{1-x}{\rm MnO}_{2}$ compound; $\mu_{\rm Li}^{\ast}$ is the chemical potential of Li which is now given in its explicit form: $\mu_{\rm Li}^{\ast} = E_{\rm tot}(\rm Li) + \mu_{\rm Li}$, with $E_{\rm tot}(\rm Li)$ being the total energy per atom of metallic Li and the chemical potentials of the Li$^{+}$ and $e^{-}$ components of the extracted Li atom both included in $\mu_{\rm Li}$. During the delithiation process, the lithium vacancies are electrochemically activated, i.e.,
\begin{equation}\label{eq;ef0}
E^f(xV_{\mathrm{Li}}^0) = 0,
\end{equation}
assuming the vacancies readily form under the influence of an external power source with an extraction voltage $V$. In addition, by assuming equilibrium with a metallic Li anode (Li/Li$^{+}$) and the external power source which acts as a reservoir of the electrons $e^{-}$, i.e., zero overpotential, the chemical potential of Li can be expressed as
\begin{equation}\label{eq;muli}
\mu_{\rm Li}^{\ast} = E_{\rm tot}({\rm Li}) - eV,
\end{equation}
where $e$ is  is the absolute value of the electron charge. 

From Eqs.~(\ref{eq:nvli})$-$(\ref{eq;muli}), the lithium-extraction voltage can be expressed in terms of the total energies as
\begin{equation}\label{eq:vol1}
V = \frac{E_{\mathrm{tot}}(xV_{\mathrm{Li}}^0) - E_{\mathrm{tot}}({\mathrm{bulk}}) + xE_{\mathrm{tot}}({\mathrm{Li}})}{xe}.
\end{equation}
This expression is applicable not only in the dilute lithium vacancy limit. In fact, $x$ can be used to describe the lithium-content difference between any two intercalation limits, and $E_{\mathrm{tot}}({\mathrm{bulk}})$ can be the total energy of any starting composition chosen as the host material. In that case, $V$ should be regarded as the average voltage between the two limits; and expression (\ref{eq:vol1}) is equivalent to that for the average voltage ({\it vs.}~Li/Li$^{+}$) previously derived by Aydinol {\it et al}.~\cite{Aydinol1997} by considering the electrical energy caused by charge displacement, assuming all due to Li. A similar expression for the voltage associated with lithiation can also be derived by regarding the lithium-insertion process as the formation of lithium interstitials.

With this formulation, one can investigate the spatial dependence of the extraction voltage, e.g., in the bulk {\it vs}.~at the surface, or explicitly calculate the voltage associated with a specific lithium-extraction mechanism. In certain mechanisms, lithium may not be the only species that is extracted during delithiation. The voltage associated with electrochemical extraction of lithium and any other species, or of any species other than lithium, can be determined in a similar way, starting from Eq.~(\ref{eq:eform}) and using appropriate thermodynamic equilibrium conditions. It is important to note that these conditions are likely to be different from those obtained by considering equilibria with all possible competing phases discussed in Sec.~\ref{sec;formenergy}. This difference is because materials synthesis, for instance, and delithiation are very distinct processes and occur in different stages of preparation and use of the material.    

\begin{figure}
\vspace{0.2cm}
\centering
\includegraphics[height=5.0cm]{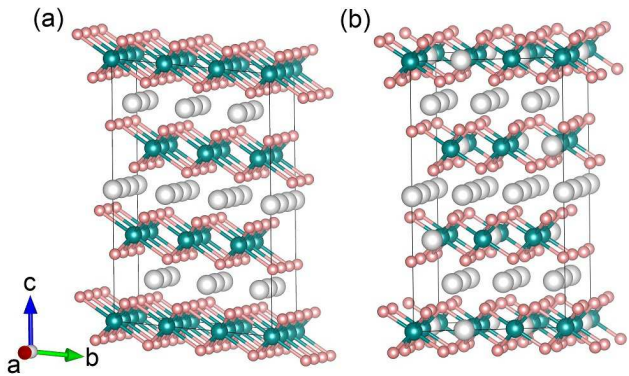}
\caption{Supercell models for (a) \ce{LiMnO2} and (b) \ce{Li2MnO3}. Large (gray) spheres are Li, medium (blue) spheres are Mn, and small (red) spheres are O. Jahn-Teller distortion is observed in \ce{LiMnO2}. The structural models shown here and in Fig.~\ref{fig;polarons} are generated by using the VESTA visualization package \cite{vesta}.}
\label{fig;struct}
\vspace{0.3cm}
\end{figure}

\section{Results}\label{sec;results}
\subsection{Bulk properties}\label{sec;bulk}

Layered \ce{LiMnO2} and \ce{Li2MnO3} are described in terms of hexagonal supercells, each containing 108 atoms and being similar to that of layered \ce{LiCoO2} and \ce{LiNiO2} \cite{Hoang2014}. Figures~\ref{fig;struct}(a) and \ref{fig;struct}(b) show these supercells after full structural optimization. For \ce{LiMnO2}, the initial supercell relaxes to a triclinically distorted structure with a cell volume of 37.11 {\AA}$^{3}$ per formula unit (f.u.), in agreement with the experimental value of 37.06 {\AA}$^{3}$ \cite{Armstrong1996}. The Mn ions in \ce{LiMnO2} are stable as high-spin Mn$^{3+}$ with a calculated magnetic moment of 3.74 $\mu_{\rm B}$. There are four short Mn$-$O bonds (1.92 {\AA}) and two long Mn$-$O bonds (2.33 {\AA}) due to the Jahn-Teller effect associated with the Mn$^{3+}$ ions. For \ce{Li2MnO3}, the supercell stays hexagonal with a cell volume of 49.35 {\AA}$^{3}$ per f.u., compared to the experimental value of 49.72 {\AA}$^{3}$ \cite{Massarotti1997}. The Mn ions in \ce{Li2MnO3} are stable as high-spin Mn$^{4+}$ with a magnetic moment of 2.98 $\mu_{\rm B}$. All six Mn$-$O bonds have a bond length of about 1.90 {\AA}. We find that an in-plane antiferromagnetic spin configuration for the manganese array gives a lower total energy than the ferromagnetic configuration, but by only 10 meV per f.u.~in the case of \ce{LiMnO2} or 4 meV per f.u.~in the case of \ce{Li2MnO3}.

\begin{figure}
\centering
\vspace{0.1cm}
\includegraphics[height=6.8cm]{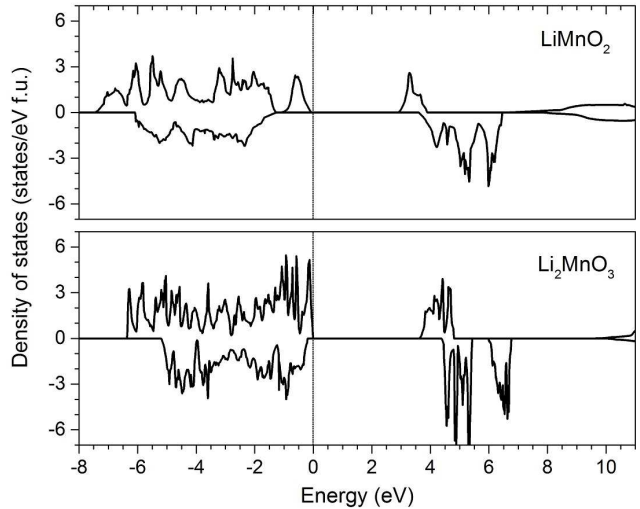}
\caption{The electronic density of states of ferromagnetic \ce{LiMnO2} and \ce{Li2MnO3}. The spin-majority spectrum is on the $+y$ axis and the spin-minority spectrum is on the $-y$ axis. The zero of energy is set to the highest occupied state.}
\label{fig;dos}
\end{figure}

The determination of the finite-cell-size correction term $\Delta^q$ in Eq.~(\ref{eq:eform}) using the Freysoldt {\it et al.}'s approach \cite{Freysoldt,Freysoldt11} requires values for the static dielectric constant, which can be obtained from DFT calculations. We find the electronic contribution to the static dielectric constant of \ce{LiMnO2} (\ce{Li2MnO3}) is 4.05 (4.75) in HSE06, based on the real part of the dielectric function $\epsilon_{1}(\omega)$ for $\omega\rightarrow0$. The ionic contribution is calculated using density-functional perturbation theory \cite{Wu2005,dielectricmethod}, within GGA$+U$ with $U=4.84$ eV for the Mn 3d states. This $U$ value is taken as an average value of those for Mn$^{3+}$ (4.64 eV) and Mn$^{4+}$ (5.04 eV) \cite{Zhou:2004p104}. The total dielectric constants are calculated to be 32.52 for \ce{LiMnO2} and 17.69 for \ce{Li2MnO3}. For comparison, the static dielectric constants of MnO, \ce{MnO2}, and \ce{Mn2O3} are 18.0$\pm$0.5, about $10^{4}$, and 8, respectively, at room temperature \cite{Young1973}. To our knowledge, the experimental static dielectric constants of \ce{LiMnO2} and \ce{Li2MnO3} are not yet available.

Figure~\ref{fig;dos} shows the total electronic density of states of layered \ce{LiMnO2} and \ce{Li2MnO3}. An analysis of the wave functions shows that the valence-band maximum (VBM) of \ce{LiMnO2} consists of 39\% from the Mn 3d states and 30\% from each O atom; the conduction-band minimum (CBM) consists of 72\% from the Mn 3d states and 12\% from each O atom. The calculated band gap is 2.90 eV. For \ce{Li2MnO3}, the VBM is predominantly O 2p states (11\%  from the Mn atom and 88\% from the three O atoms) and the CBM is predominantly Mn 3d states (68\% from the Mn atom and 28\% from the O atoms). The calculated band gap is 3.62 eV. In both compounds, the Li 2s state is high up in the conduction band, indicating that Li donates its electron to the lattice and becomes Li$^{+}$. \ce{LiMnO2} (\ce{Li2MnO3}) can be thus be regarded nominally as an ordered arrangement of Li$^{+}$, Mn$^{3+}$ (Mn$^{4+}$), and O$^{2-}$ units. As will be illustrated in Secs.~\ref{sec;formation} and \ref{sec;migration}, the formation and migration of intrinsic point defects in the materials are directly related to their structural and electronic properties, especially the nature of the electronic states at the VBM and CBM.

\begin{figure}[t!]
\centering
\includegraphics[height=7.6cm]{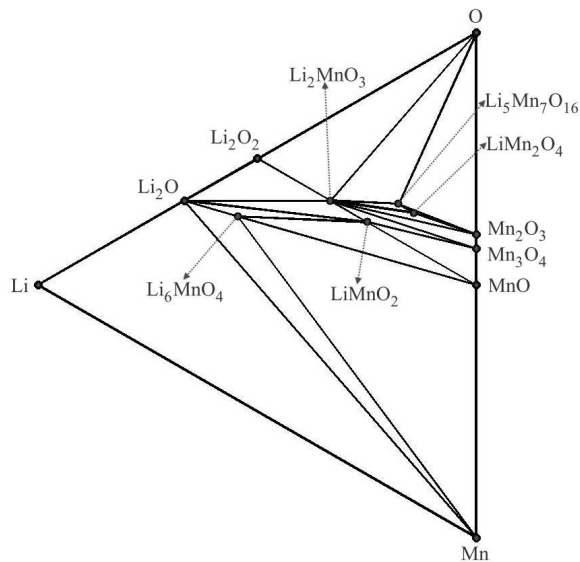}
\caption{Calculated ternary phase diagram for the Li-Mn-O system at 0 K; only the stable compounds are shown. The phase diagram shown here is generated by using a MATLAB application developed by Ong \cite{Ong2008}.}
\label{fig;phase}
\end{figure}

\begin{table}
\small
\caption{Calculated formation enthalpies at 0 K, in eV per formula unit. Experimental values at 298 K are also included.}\label{tbl;enthalpies}
\begin{center}
\begin{ruledtabular}
\begin{tabular}{llrr}
Compound & Crystal structure &This work & Experiments \\
\colrule
\ce{Li2O} & cubic & $-$5.75 & $-$6.21 (Ref.~\cite{chase}) \\
\ce{Li2O2} & hexagonal & $-$5.84 &$-$6.56 (Ref.~\cite{chase}) \\
\ce{MnO} & cubic & $-$4.16 & $-$3.96 (Ref.~\cite{Knacke1991}) \\
\ce{MnO2} & tetragonal & $-$4.98 & $-$5.41 (Ref.~\cite{Knacke1991}) \\
\ce{Mn2O3} & orthorhombic & $-$10.09 & $-$9.94 (Ref.~\cite{Knacke1991})\\
\ce{Mn3O4}& tetragonal & $-$14.67 & $-$14.37 (Ref.~\cite{Knacke1991}) \\
\ce{LiMnO2} & monoclinic & $-$8.43 & $-$8.59 (Ref.~\cite{Wang20051230})\\
\ce{LiMn2O4} & tetragonal & $-$13.89 &$-$14.31 (Ref.~\cite{Wang20051182})\\
\ce{Li2MnO2} & trigonal & $-$9.94 & \\
\ce{Li2MnO3} & monoclinic & $-$12.30 &\\
\ce{Li2Mn3O7} & triclinic & $-$22.47 &\\
\ce{Li4Mn5O12} & monoclinic & $-$40.33 &\\
\ce{Li5Mn7O16} & orthorhombic & $-$54.39 &\\
\ce{Li6MnO4} & tetragonal & $-$21.55 &  \\
\end{tabular}
\end{ruledtabular}
\end{center}
\end{table}

\subsection{Phase diagram and chemical potentials}\label{sec;phasediagram}

\begin{figure*}[t!]
\centering
\hspace{0.2cm}
\includegraphics[height=6.5cm]{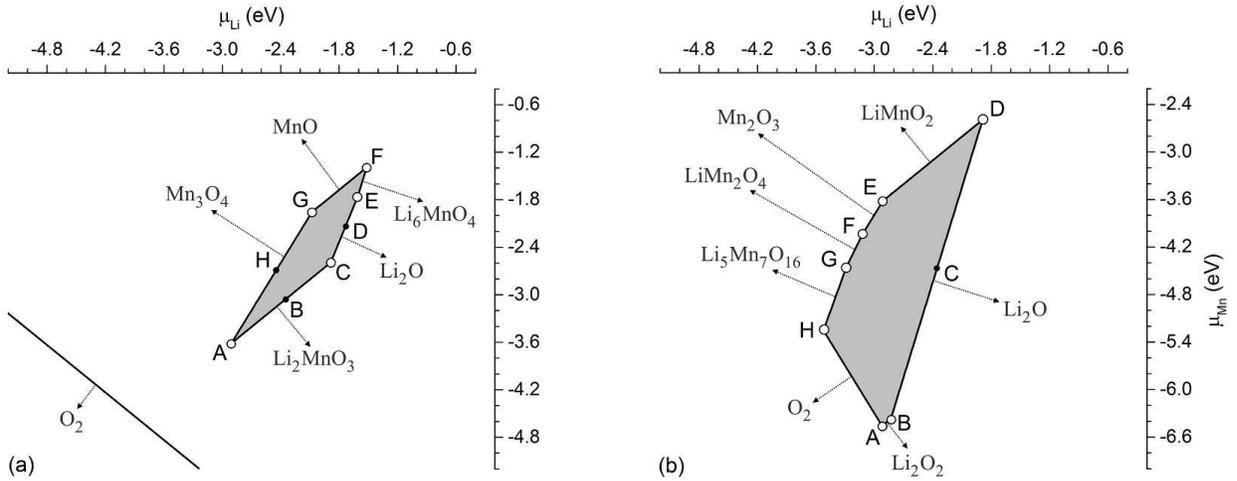}
\caption{Chemical-potential diagrams for (a) \ce{LiMnO2} and (b) \ce{Li2MnO3}. Only phases that define the stability region of the compounds, here shown as a shaded polygon, are included. In (a), \ce{O2} is also included for reference; in (b), \ce{Mn3O4} is present but not visible since it is very close to point E. The large white dots represent three-phase equilibria; the small black dots are midpoints between the white dots.}
\label{fig;chempot}
\end{figure*}

Figure~\ref{fig;phase} shows the phase diagram for the Li-Mn-O ternary system at 0 K, constructed using the calculated formation enthalpies listed in Table~\ref{tbl;enthalpies} and a phase-diagram construction method based on the convex hull approach \cite{Wang2007,Ong2008}. The listed compounds are taken from those Li-Mn-O phases available in the Materials Project database~\cite{Jain2013} and calculated using the HSE06 functional. We note that, in the formation enthalpy calculations, different crystal structures and manganese spin configurations and charge states are investigated and only the lowest energy configurations are reported. The formation enthalpies of \ce{LiMn2O4}, \ce{Mn2O3}, \ce{Li2MnO3}, and \ce{Li5Mn7O16} were already reported in Ref.~\cite{HoangLiMn2O4}. but are also included here in Table~\ref{tbl;enthalpies} for completeness. The phase diagram shows equilibria between \ce{LiMnO2} and competing Li-Mn-O phases such as \ce{Li2O}, \ce{Li2MnO3}, \ce{Mn3O4}, \ce{MnO}, and \ce{Li6MnO4}, and between \ce{Li2MnO3} and \ce{O2}, \ce{Li5Mn7O16}, \ce{LiMn2O4}, \ce{Mn2O3}, \ce{Mn3O4}, \ce{LiMnO2}, \ce{Li2O}, and \ce{Li2O2}. These competing phases ultimately define the range of the atomic chemical potential values, shown as a shaded polygon in Fig.~\ref{fig;chempot}(a) or \ref{fig;chempot}(b), in which the host compound is stable. Points A$-$H in Figs.~\ref{fig;chempot}(a) and \ref{fig;chempot}(b) represent three-phase equilibria associated with \ce{LiMnO2} and \ce{Li2MnO3}, respectively, or midpoints between two three-phase equilibria. We note that layered \ce{LiMnO2} would be unstable toward competing Li-Mn-O phases if in the calculations the Jahn-Teller distortion were not allowed. Besides, layered and orthorhombic \ce{LiMnO2} phases are degenerate at 0 K; the energy difference is within 1 meV.   

\subsection{Defect structure and energetics}\label{sec;formation}

\begin{figure*}
\centering
\includegraphics[height=7.0cm]{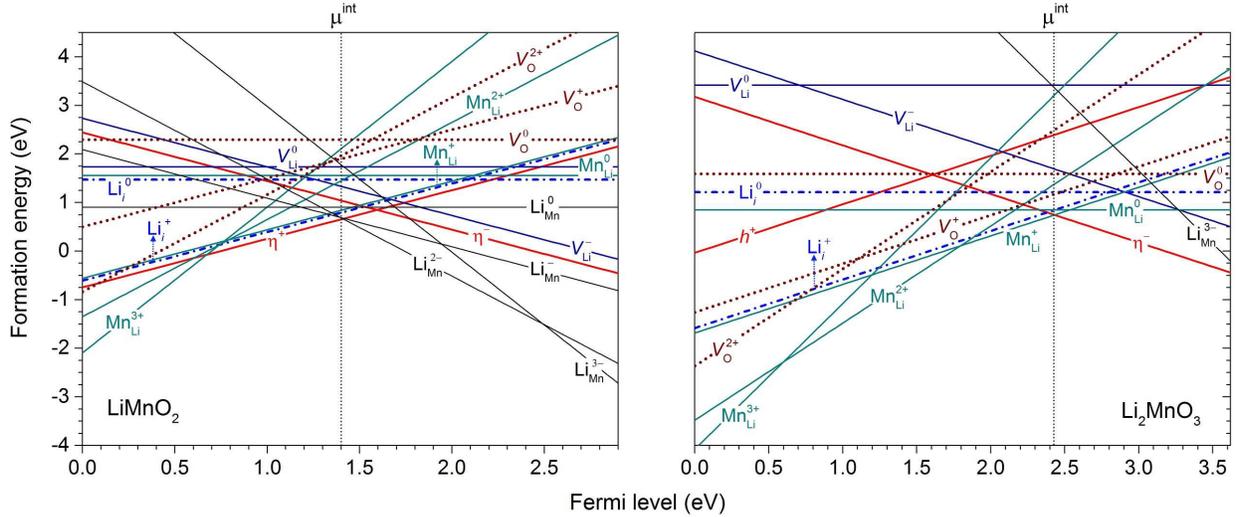}
\caption{Calculated formation energies of intrinsic point defects in \ce{LiMnO2} and \ce{Li2MnO3}, plotted as a function of the Fermi level with respect to the VBM. The energies in \ce{LiMnO2} are obtained at point C in Fig.~\ref{fig;chempot}(a) and those in \ce{Li2MnO3} are at point D in Fig.~\ref{fig;chempot}(b). In the chemical-potential diagrams, these two points correspond to thermodynamic equilibrium between \ce{LiMnO2}, \ce{Li2MnO3}, and \ce{Li2O}. In the absence of extrinsic charged impurities, the Fermi level of the system is at $\mu_{e}=\mu_{e}^{\rm{int}}$, where charge neutrality is maintained.}
\label{fig;formenergy}
\end{figure*}

Figure~\ref{fig;formenergy} shows the calculated formation energies of low-energy defects in layered \ce{LiMnO2} and \ce{Li2MnO3}, obtained at points C and D in Figs.~\ref{fig;chempot}(a) and \ref{fig;chempot}(b), respectively. These defects include delocalized electron hole (hereafter denoted as $h^{+}$); hole ($\eta^{+}$) and electron ($\eta^{-}$) polarons; lithium vacancies ($V_{\rm Li}$), interstitials (Li$_{i}$), and antisites (Li$_{\rm Mn}$); manganese vacancies ($V_{\rm Mn}$) and antisites (Mn$_{\rm Li}$); and oxygen vacancies ($V_{\rm O}$) in different charge states. For each defect, calculations are carried out in various configurations (i.e., at different lattice sites and/or in different spin states); however, only the lowest-energy configurations of the defects are reported, unless otherwise noted. In the absence of electrically active impurities that can shift the Fermi-level position or when such impurities occur in much lower concentrations than charged intrinsic defects, the Fermi level is at $\mu_{e}^{\rm int}$, determined only by the intrinsic defects. With the chosen set of the atomic chemical potentials, the Fermi level of \ce{LiMnO2} is at $\mu_{e}^{\rm int}=1.40$ eV, predominantly defined by hole polarons ($\eta^{+}$) and negatively charged lithium antisites (Li$_{\rm Mn}^{2-}$); {\it cf.}~Fig.~\ref{fig;formenergy}; for \ce{Li2MnO3}, $\mu_{e}^{\rm int}$ is at 2.43 eV, almost exclusively defined by electron polarons ($\eta^{-}$) and positively charged manganese antisites (Mn$_{\rm Li}^{+}$). 

\begin{table*}
\caption{Calculated formation energies ($E^{f}$) and binding energies ($E_{b}$) of relevant point defects in \ce{LiMnO2} and \ce{Li2MnO3}, obtained at points A$-$H in Figs.~\ref{fig;chempot}(a) and \ref{fig;chempot}(b). The manganese ion associated with each elementary defect is listed in parentheses.}\label{tab:formenergy}
\begin{center}
\begin{ruledtabular}
\begin{tabular}{lllllllllllr}
&&\multicolumn{8}{c}{$E^{f}$ (eV)}& \\
\colrule
&Defect&A&B&C&D&E&F&G&H&Constituents&$E_{b}$ (eV) \\
\colrule
\ce{LiMnO2}&$\eta^{+}$&0.48&0.66&0.66&0.78&0.97&1.15&1.15&0.81&(Mn$^{4+}$)\\ 
&$\eta^{-}$ &1.21&1.04&1.04&0.92&0.73&0.55&0.55&0.89&(Mn$^{2+}$)\\ 
&$V_{\rm{Li}}^-$&0.48&0.82&1.33&1.35&1.30&1.21&0.65&0.58&\\  
&$V_{\rm{Li}}^0$&0.71&1.22&1.73&1.87&2.01&2.10&1.54&1.13&$V_{\rm{Li}}^- + \eta^{+}$&0.25\\
&Li$_{i}^+$&1.65&1.32&0.80&0.79&0.84&0.93&1.49&1.56&\\ 
&Li$_{i}^0$&2.49&1.98&1.47&1.33&1.19&1.10&1.66&2.08&Li$_{i}^+ + \eta^{-}$&0.37\\ 
&Li$_{\rm{Mn}}^{2-}$&1.03&0.68&0.68&0.71&0.62&0.53&0.53&0.80&\\  
&Li$_{\rm{Mn}}^-$&0.86&0.69&0.69&0.84&0.93&1.03&1.03&0.95&Li$_{\rm{Mn}}^{2-} + \eta^{+}$&0.65\\
&Li$_{\rm{Mn}}^0$&0.90&0.90&0.90&1.18&1.45&1.73&1.73&1.32&Li$_{\rm{Mn}}^{2-} + 2\eta^{+}$&1.09\\
&Li$_{\rm{Mn}}^{3-}$&2.29&1.77&1.77&1.67&1.40&1.13&1.13&1.74&Li$_{\rm{Mn}}^{2-} + \eta^{-}$&$-$0.05\\
&Mn$_{\rm{Li}}^+$&0.67&0.84&0.84&0.69&0.60&0.50&0.50&0.58& (Mn$^{2+}$)\\ 
&Mn$_{\rm{Li}}^0$&1.56&1.56&1.56&1.28&1.00&0.73&0.73&1.14&Mn$_{\rm{Li}}^+ + \eta^{-}$&0.33\\
&Mn$_{\rm{Li}\ast}^{2+}$&1.11&1.45&1.45&1.43&1.52&1.61&1.61&1.34&Mn$_{\rm{Li}}^+ + \eta^{+}$&0.04\\ 
&Mn$_{\rm{Li}}^{3+}$&1.60&2.11&2.11&2.21&2.49&2.76&2.76&2.15&Mn$_{\rm{Li}}^{2+} + \eta^{+}$&\\
&Mn$_{\rm{Li}}^{+}$-$V_{\rm{Li}}^{-}$&0.84&1.35&1.86&1.72&1.58&1.40&0.84&0.84&Mn$_{\rm{Li}}^{+} + V_{\rm{Li}}^{-}$&0.32\\
&Mn$_{\rm{Li}}$-${\rm Li}_{\rm{Mn}}$&1.03&1.03&1.03&1.03&1.03&1.03&1.03&1.03&Mn$_{\rm{Li}}^{+} + \rm{Li}_{\rm{Mn}}^{2-} + \eta^{+}$&1.15\\
&$V_{\rm{O}}^{2+}$&2.64&2.48&1.97&1.94&2.03&2.17&2.73&2.66&\\  
&$V_{\rm{O}}^{+}$&2.75&2.41&1.90&1.75&1.66&1.61&2.17&2.45&$V_{\rm{O}}^{2+} + \eta^{-}$&1.11\\
&$V_{\rm{O}}^{0}$&3.32&2.81&2.29&2.02&1.74&1.51&2.07&2.69&$V_{\rm{O}}^{2+} + \eta^{-} + \eta_{\ast}^{-}$&\\
\ce{Li2MnO3}&$h^{+}$&1.78&1.86&2.11&2.40&2.36&2.15&1.94&1.62&\\ 
&$\eta^{-}$ &1.36&1.28&1.03&0.75&0.79&0.99&1.20&1.52&(Mn$^{3+}$)\\ 
&$V_{\rm{Li}}^-$&1.27&1.28&1.51&1.69&0.71&0.71&0.75&0.83&\\ 
&$V_{\rm{Li}}^0$&2.38&2.47&2.94&3.42&2.39&2.18&2.02&1.78&$V_{\rm{Li}}^- + \eta_{\rm O}^{+}$&\\
&Li$_{i}^+$&1.27&1.26&1.03&0.85&1.83&1.83&1.79&1.70&\\ 
&Li$_{i}^0$&2.24&2.16&1.68&1.21&2.24&2.45&2.61&2.85&Li$_{i}^+ + \eta^{-}$&0.38\\
&Li$_{\rm{Mn}}^{3-}$&2.36&2.12&2.80&3.36&3.48&3.90&4.27&4.67&\\ 
&Li$_{\rm{Mn}}^{2-}$&2.18&2.02&2.94&3.79&3.87&4.08&4.24&4.32&Li$_{\rm{Mn}}^{3-} + \eta_{\rm O}^{+}$&\\ 
&Li$_{\rm{Mn}}^-$&2.40&2.32&3.49&4.62&4.66&4.66&4.62&4.37&Li$_{\rm{Mn}}^{3-} + 2\eta_{\rm O}^{+}$&\\
&Li$_{\rm{Mn}}^0$&3.27&3.27&4.69&6.10&6.10&5.89&5.64&5.08&Li$_{\rm{Mn}}^{3-} + 2\eta_{\rm O}^{+} + \eta_{\rm O*}^{+}$&\\
&Mn$_{\rm{Li}}^+$&2.97&3.05&1.88&0.75&0.71&0.71&0.75&0.99&(Mn$^{2+}$)\\ 
&Mn$_{\rm{Li}}^0$&3.68&3.68&2.27&0.85&0.85&1.06&1.31&1.87&Mn$_{\rm{Li}}^{2+} + 2\eta^{-}$&2.02\\
&Mn$_{\rm{Li}}^{2+}$&2.99&3.15&2.23&1.38&1.30&1.09&0.93&0.85&(Mn$^{3+}$)&\\ 
&Mn$_{\rm{Li}}^{3+}$&4.17&4.41&3.72&3.17&3.05&2.63&2.25&1.86&(Mn$^{4+}$)\\
&Mn$_{\rm{Li}}^{+}$-$V_{\rm{Li}}^{-}$&3.84&3.93&2.98&2.04&1.01&1.01&1.10&1.43&Mn$_{\rm{Li}}^{+} + V_{\rm{Li}}^{-}$&0.40\\
&Mn$_{\rm{Li}}$-Li$_{\rm{Mn}}$&1.96&1.96&1.96&1.96&1.96&1.96&1.96&1.96&Mn$_{\rm{Li}}^{3+} + \rm{Li}_{\rm{Mn}}^{3-}$&4.56\\
&$V_{\rm{O}}^{2+}$&3.25&3.32&2.87&2.50&3.44&3.30&3.13&2.92&\\  
&$V_{\rm{O}}^{+}$&2.54&2.53&1.83&1.17&2.16&2.23&2.27&2.37&$V_{\rm{O}}^{2+} + \eta^{-}$&2.07\\
&$V_{\rm{O}}^{0}$&3.57&3.48&2.54&1.59&2.61&2.89&3.15&3.57&$V_{\rm{O}}^{2+} + \eta^{-} + \eta_{\ast}^{-}$&\\
\end{tabular}
\end{ruledtabular}
\end{center}
\end{table*}

Since defect-formation energies are functions of the atomic chemical potentials which represent the experimental conditions under which the defects are created, the results presented in Fig.~\ref{fig;formenergy} are not the only scenario that may occur. We list in Table~\ref{tab:formenergy} the calculated formation energies of relevant intrinsic point defects under conditions at points A$-$H in the chemical-potential diagrams. The allowed range of the oxygen chemical potential, $\mu_{\rm O}$, is from $-$0.96 [point A in Fig.~\ref{fig;chempot}(a)] to $-$2.76 eV [point F in Fig.~\ref{fig;chempot}(a)] in \ce{LiMnO2}, or from 0 [the A$-$H line in Fig.~\ref{fig;chempot}(b)] to $-$1.98 eV [point D in Fig.~\ref{fig;chempot}(b)] in \ce{Li2MnO3}. The oxygen chemical potential can be controlled by controlling temperature and pressure [{\it cf.}~Eq.~(\ref{eq;muO})] and/or oxygen-reducing agents. Lower $\mu_{\rm O}$ values are usually associated with higher temperatures and/or lower oxygen partial pressures and/or the presence of oxygen-reducing agents. \ce{Li2MnO3}, for example, is usually prepared by solid-state reaction in the temperature range from 500$^\circ$C to 950$^\circ$C \cite{Sathiya2013,Massarotti1997,Rossouw1991,Kalyani1999,Kubota2012,Robertson2002}. If one assumes an oxygen partial pressure of 0.2 atm and no oxygen-reducing agents, this temperature range gives $\mu_{\rm O}$ values in the range from $-$0.87 eV to $-$1.47 eV \cite{Reuter2001}, which correspond to conditions approximately within the region enclosed by points C (where $\mu_{\rm O} = -$1.03 eV), D ($-$1.98 eV), and E ($-$0.95 eV) in Fig.~\ref{fig;chempot}(b). In the presence of oxygen-reducing agents, e.g., \ce{CaH2} or LiH, $\mu_{\rm O}$ is expected to take a very low value even at low temperatures, e.g., 255$-$265$^\circ$C \cite{Kubota2012}. For each set of the chemical potentials, the formation-energy values reported in Table~\ref{tab:formenergy} are obtained at the respective Fermi-level position $\mu_{e}^{\rm int}$, determined by the charge-neutrality condition [{\it cf.}~Eq.~(\ref{eq:neutrality})]. We find that $\mu_{e}^{\rm int}$ is at 1.23$-$1.89 eV in \ce{LiMnO2} or 1.65$-$2.43 eV in \ce{Li2MnO3}, which is always away from both the VBM and CBM. Overall, we find that certain point defects in the two compounds have very low formation energies and, hence, can occur in the materials with high concentrations, e.g., during synthesis. These defects, except the mobile ones, are expected to get trapped when the material is cooled to room temperature. We also find that many of the charged defects have positive formation energies only in a small region near midgap. Before discussing the implications of these findings, let us describe in detail the structure and energetics of the defects.

{\bf Electronic defects.} Let us first examine those defects that are created by removing (adding) an electron from (to) the bulk supercells. The removal of an electron from layered \ce{LiMnO2} results in the formation of a high-spin Mn$^{4+}$ ion with a calculated magnetic moment of 3.10 $\mu_{\rm B}$, i.e., a localized electron hole, at one of the Mn$^{3+}$ sites. The lattice geometry near Mn$^{4+}$ is distorted with respect to the perfect bulk lattice with the six neighboring O atoms moving toward the Mn$^{4+}$, resulting in four Mn$-$O bonds with a bond length of 1.90 {\AA} and two slightly longer Mn$-$O bonds with a bond length of 1.96 {\AA}. This local lattice distortion and the localized hole constitute a quasiparticle called {\it hole polaron}, hereafter denoted as $\eta^{+}$, in which the hole is self-trapped in its own potential. The addition of an electron, on the other hand, leads to the creation of a high-spin Mn$^{2+}$ ion with a magnetic moment of 4.51 $\mu_{\rm B}$, i.e., a localized electron, at one of the Mn$^{3+}$ sites. The lattice geometry near Mn$^{2+}$ is also distorted as compared to the perfect bulk compound; there are four Mn$-$O bonds with a bond length of 2.07 {\AA} and two Mn$-$O bonds with the average bond length of 2.33 {\AA}. The electronic defect associated with this localized electron is called an {\it electron polaron}, denoted as $\eta^{-}$. As expected, the Jahn-Teller distortion almost vanishes at the $\eta^{+}$ and $\eta^{-}$ sites because Mn$^{4+}$ and Mn$^{2+}$ ions are not Jahn-Teller active. The calculated formation energy of $\eta^{+}$ ($\eta^{-}$) is found to be 0.48$-$1.15 eV (0.55$-$1.21 eV), depending on the specific set of the atomic chemical potentials; {\it cf.}~Table \ref{tab:formenergy}. 

In \ce{Li2MnO3}, the removal of an electron results in an electron hole, denoted as $h^{+}$, that is delocalized all over the oxygen sites in the supercell. The addition of an electron to the supercell, on the other hand, leads to the creation of a high-spin Mn$^{3+}$ with a calculated magnetic moment of 3.70 $\mu_{\rm B}$, i.e., a localized electron, at one of the Mn$^{4+}$ sites. At the Mn$^{3+}$ site, there are four short Mn$-$O bonds with an average bond length of 1.93 {\AA} and two long Mn$-$O bonds with a bond length of 2.10 {\AA}. The presence of the Mn$^{3+}$ ion thus introduces local Jahn-Teller distortion into \ce{Li2MnO3}. This localized electron and the local lattice distortion constitute an electron polaron, hereafter also denoted as $\eta^{-}$ (one, however, should not be confused with $\eta^{-}$ in \ce{LiMnO2} which is associated with Mn$^{2+}$). The formation energy of $h^{+}$ ($\eta^{-}$) is 1.62$-$2.40 eV (0.75$-$1.52 eV), depending on the chemical potentials; {\it cf.}~Table \ref{tab:formenergy}.

As the electron removal (addition) process occurs at the VBM (CBM), the formation of the electronic defects in \ce{LiMnO2} or \ce{Li2MnO3} is directly related to the electronic states at the band edges. For example, unbound hole polarons cannot be stabilized in \ce{Li2MnO3} because the VBM of the material is predominantly O 2p states, unlike in \ce{LiMnO2} where the majority of the electronic states at the VBM are Mn 3d states. The self-trapping energies of unbound $\eta^{+}$ and $\eta^{-}$ in \ce{LiMnO2} are 0.70 and 0.55 eV, respectively, defined as the difference between the formation energy of the free hole or electron and that of the hole or electron polaron \cite{Hoang2014}. In \ce{Li2MnO3}, the self-trapping energy of $h^{+}$ is, of course, 0 eV, and that of unbound $\eta^{-}$ is 0.47 eV. Finally, since the lattice distortion associated with the polarons $\eta^{+}$ and $\eta^{-}$ is limited mainly to their neighboring O atoms, they can be regarded as {\it small polarons} \cite{Stoneham2007}.

{\bf Vacancies and interstitials.} Let us now examine those defects whose formation involves the exchange of ions (and electrons) with reservoirs. The creation of $V_{\rm Li}^-$ in \ce{LiMnO2} involves removing a Li$^{+}$ ion, which causes negligible disturbance in the local lattice environment. $V_{\rm Li}^0$ is, on the other hand, created by removing a Li atom, which is in fact a Li$^{+}$ ion and an electron from a neighboring Mn atom. This removal results in a void at the site of the removed Li$^{+}$, i.e., $V_{\rm Li}^-$, and a high-spin Mn$^{4+}$, i.e., $\eta^{+}$, at the neighboring Mn site. $V_{\rm Li}^0$ is thus a complex of $V_{\rm Li}^-$ and $\eta^{+}$; it has a binding energy of 0.25 eV with respect to its constituents. The formation energy of the lithium vacancies is found to be 0.48$-$1.35 eV, depending on the chemical potentials. For the lithium interstitials, Li$_{i}^{+}$ is created by adding a Li$^{+}$. This defect is found to reside in the Li layer. Because of the repulsive Coulomb interaction between this and other Li$^{+}$ ions, there is significant rearrangement of the Li$^{+}$ ions in the Li layer. Li$_{i}^{0}$, created by adding a Li atom, is a complex of Li$^{+}$ and $\eta^{-}$ with a binding energy of 0.37 eV. The formation energy of the lithium interstitials is 0.79$-$1.65 eV. We note that defects such as $\eta^{+}$, $\eta^{-}$, $V_{\rm Li}^-$, and Li$_{i}^{+}$ are regarded as {\it elementary defects}; other defects, e.g., $V_{\rm Li}^0$ and Li$_{i}^{0}$, can be interpreted in terms of these basic building blocks.

\begin{figure}
\vspace{0.1cm}
\centering
\includegraphics[height=4.0cm]{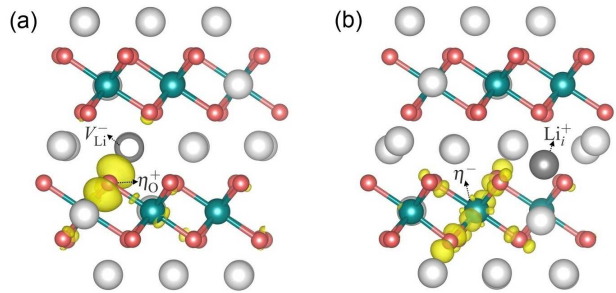}
\caption{Structures of (a) $V_{\rm Li}^{0}$ and (b) Li$_{i}^{0}$ in \ce{Li2MnO3}. $V_{\rm Li}^{0}$ is a defect complex of $V_{\rm Li}^{-}$ (large hollow sphere) and $\eta_{\rm O}^{+}$, i.e., O$^{-}$ at the O$^{2-}$ site, whereas Li$_{i}^{0}$ is a complex of Li$_{i}^{+}$ (large dark sphere) and $\eta^{-}$, i.e., Mn$^{3+}$ at the Mn$^{4+}$ site. Charge densities associated with the oxygen hole polaron $\eta_{\rm O}^{+}$ and manganese electron polaron $\eta^{-}$ are visualized as isosurfaces.}
\label{fig;polarons}
\end{figure}

In \ce{Li2MnO3}, lithium vacancies and interstitials are created in similar ways. We find that the creation of $V_{\rm Li}^-$ at the 4$h$ and 2$c$ sites of the space group $C2/m$, i.e., in the Li layer, is energetically more favorable than at the 2$b$ site, i.e., in the Mn/Li layer; the formation energies are lower by 0.26 and 0.24 eV, respectively. The removal of a Li atom, i.e., the creation of $V_{\rm Li}^0$, results in a void left by the removed atom, i.e., $V_{\rm Li}^-$, and a localized hole at a neighboring oxygen site, hereafter denoted as $\eta_{\rm O}^{+}$. $V_{\rm Li}^0$ is therefore a complex of $V_{\rm Li}^-$ and $\eta_{\rm O}^{+}$, see Fig.~\ref{fig;polarons}(a). $\eta_{\rm O}^{+}$ has a calculated magnetic moment of $-$0.69 $\mu_{\rm B}$, as compared to that of approximately 0 $\mu_{\rm B}$ at the other oxygen sites. The minus sign in the magnetic moments indicates that the interaction between Mn and O is antiferromagnetic. Our results thus indicate that, in the presence of $V_{\rm Li}^-$, one of the O$^{2-}$ ions becomes O$^{-}$. The neighboring Li$^{+}$ and Mn$^{4+}$ ions of O$^{-}$ move slightly away from the site as the negative charge gets reduced. The Mn$-$O bonds associated with $\eta_{\rm O}^{+}$ have a bond length of 1.94 {\AA}, compared to 1.90 {\AA} of the other Mn$-$O bonds. With this local lattice distortion, $\eta_{\rm O}^{+}$ can be referred to as {\it oxygen hole polaron}, also called ``O$^{-}$ bound polaron'' \cite{Schirmer2006}. We note that this $\eta_{\rm O}^{+}$ is about 0.1 eV lower in energy than a different configuration where the hole is localized on two neighboring O atoms of $V_{\rm Li}^-$, hereafter denoted as $\eta_{\rm O*}^{+}$. We further note that $\eta_{\rm O}^{+}$ is stable only in the presence of $V_{\rm Li}^-$ or, in principle, some other negatively charged defect such as a lithium antisite or manganese vacancy inside the material. The formation energy of the lithium vacancies is 0.71$-$1.69 eV. For the lithium interstitials, Li$_{i}^{+}$ is most stable in the space between the Mn/Li layer and the Li layer. There is significant rearrangement of the Li$^{+}$ ions due to repulsive Coulomb interaction. Li$_{i}^{0}$ is a complex of Li$_{i}^{+}$ and $\eta^{-}$, see Fig.~\ref{fig;polarons}(b), with a binding energy of 0.38 eV. The formation energy of the lithium interstitials is found to be 0.85$-$1.83 eV.

To check the robustness of our results for $V_{\rm Li}^0$, we carry out similar calculations using the GGA$+U$ method \cite{dudarev1998,liechtenstein1995} with the on-site Hubbard corrections applied to both Mn 3d and O 2p states; $U$\{Mn d\} = 4.84 eV as mentioned earlier, and $U$\{O p\} varies from 0 to 7.0 eV. For 0 $\leq$ $U$\{\rm O p\} $<$ 5.0 eV, we find that the hole in the $V_{\rm Li}^0$ complex is localized on two neighboring oxygen sites of $V_{\rm Li}^-$, which is similar to  $\eta_{\rm O*}^{+}$ mentioned above; whereas for $U$\{\rm O p\} $\geq$ 5.0 eV, the hole is localized on a single oxygen site, i.e., $\eta_{\rm O}^{+}$. Our results thus suggest that the inclusion of only the local repulsion between the Mn 3d electrons, i.e., $U$\{O p\} = 0 eV, may not be adequate in the case of \ce{Li2MnO3} where the local interaction between p electrons is also important. Besides, the physics may depend subtly on the interaction between the transition-metal (Mn) d and ligand (O) p electrons. It is therefore important to treat all the orbitals on the equal footing as in our current calculations using the HSE06 screened hybrid density functional. It should be noted that $U$\{\rm O p\} $>$ 5.0 eV is employed by other research groups to correctly capture localized defect states in doped or defective oxides \cite{Nolan2006,Morgan2009,Keating2012}. As mentioned in Ref.~\cite{Nolan2006} and references therein, the on-site Coulomb interaction for O 2p holes in oxide materials determined from experimental data is also about 5$-$7 eV.

Among the manganese vacancies in \ce{LiMnO2}, $V_{\rm Mn}^{3-}$, i.e., the removal of a Mn$^{3+}$ ion, is an elementary defect. Other defects such as $V_{\rm Mn}^{2-}$, $V_{\rm Mn}^{-}$, or $V_{\rm Mn}^{0}$ are complexes of $V_{\rm Mn}^{3-}$ and, respectively, one, two, or three $\eta^{+}$. The calculated formation energy of the manganese vacancies is found to be 1.82$-$3.05 eV, depending on the specific set of the atomic chemical potentials. For the oxygen vacancies, $V_{\rm O}^{2+}$ is an elementary defect; $V_{\rm O}^{+}$ is a defect complex of $V_{\rm O}^{2+}$ and $\eta^{-}$. The oxygen vacancies have a formation energy of 1.51$-$2.64 eV. In \ce{Li2MnO3}, $V_{\rm Mn}^{4-}$, i.e., the removal of a Mn$^{4+}$ ion, is an elementary defect. Other manganese vacancies such as $V_{\rm Mn}^{3-}$, $V_{\rm Mn}^{2-}$, or $V_{\rm Mn}^{-}$ are complexes of $V_{\rm Mn}^{4-}$ and $\eta_{\rm O}^{+}$; $V_{\rm Mn}^{0}$ is a complex of a $V_{\rm Mn}^{4-}$, two $\eta_{\rm O}^{+}$, and an $\eta_{\rm O*}^{+}$. The formation energy of these vacancies is 4.27$-$6.94 eV (not included in Table~\ref{tab:formenergy}). For the oxygen vacancies in \ce{Li2MnO3}, $V_{\rm O}^{2+}$ is an elementary defect; $V_{\rm O}^{+}$ is a defect complex of $V_{\rm O}^{2+}$ and $\eta^{-}$. These oxygen vacancies are most stable at the 8$j$ site (of the space group $C2/m$) and have a formation energy of 1.17$-$2.54 eV; the energy at the 4$i$ site is higher by 0.25$-$0.32 eV. In both compounds, $V_{\rm O}^{0}$ can be identified as a defect complex of $V_{\rm O}^{2+}$, $\eta^{-}$, and $\eta_{\ast}^{-}$, where $\eta_{\ast}^{-}$ is an electron localized at the void formed by the moved O$^{2-}$ ion. 

{\bf Antisite defects.} Lithium antisites Li$_{\rm Mn}$ are created by replacing Mn at a Mn site with Li. In \ce{LiMnO2}, Li$_{\rm Mn}^{2-}$, i.e., Li$^{+}$ substituting Mn$^{3+}$, is an elementary defect. Other antisites such as Li$_{\rm Mn}^{-}$ or Li$_{\rm Mn}^{0}$ are complexes of Li$_{\rm Mn}^{2-}$ and $\eta^{+}$. Manganese antisites Mn$_{\rm Li}$ are created by replacing Li at a Li site with Mn. We find that Mn$_{\rm Li}^{+}$ is an elementary defect. In this defect configuration, manganese is most stable as high-spin Mn$^{2+}$ with the calculated magnetic moment of 4.51 $\mu_{\rm B}$. Other defects such as Mn$_{\rm Li}^{0}$ or Mn$_{{\rm Li}\ast}^{2+}$ are defect complexes of Mn$_{\rm Li}^{+}$ and, respectively, $\eta^{-}$ or $\eta^{+}$; Mn$_{\rm Li}^{3+}$ is a defect complex of Mn$_{\rm Li}^{2+}$ (i.e., Mn$^{3+}$ replacing Li$^{+}$; not to be confused with Mn$_{{\rm Li}\ast}^{2+}$) and $\eta^{+}$. Lithium and manganese antisites in \ce{LiMnO2} have a very low formation energies, only 0.53$-$0.86 eV (Li$_{\rm Mn}$) or 0.50$-$0.84 eV (Mn$_{\rm Li}$).

In \ce{Li2MnO3}, Li$_{\rm Mn}^{3-}$, i.e., Li$^{+}$ substituting Mn$^{4+}$, is an elementary defect. Other lithium antisites such as Li$_{\rm Mn}^{2-}$ or Li$_{\rm Mn}^{-}$ are complexes of Li$_{\rm Mn}^{3-}$ and $\eta_{\rm O}^{+}$; Li$_{\rm Mn}^{0}$ is a complex of a Li$_{\rm Mn}^{3-}$, two $\eta_{\rm O}^{+}$, and an $\eta_{\rm O*}^{+}$. These defects have a formation energy of 2.02$-$4.32 eV, depending on the specific set of the atomic chemical potentials. Regarding manganese antisites Mn$_{\rm Li}$, the lowest-energy configuration is Mn$_{\rm Li}^{+}$ in which Li$^{+}$ in the Li layer is substituted by high-spin Mn$^{2+}$. This defect has a calculated formation energy of 0.71$-$3.05 eV, depending on the chemical potentials. Other manganese antisites include Mn$_{\rm Li}^{2+}$ (i.e., Mn$^{3+}$ replacing Li$^{+}$), Mn$_{\rm Li}^{0}$ (a complex of Mn$_{\rm Li}^{2+}$ and two $\eta^{-}$), and Mn$_{\rm Li}^{3+}$ (Mn$^{4+}$ replacing Li$^{+}$). Obviously manganese at the Li site can, in principle, be stable in three different charge states; and Mn$_{\rm Li}^{+}$, Mn$_{\rm Li}^{2+}$, and Mn$_{\rm Li}^{3+}$ are all elementary defects. We also find that manganese antisites are energetically more favorable in the Li layer than in the Mn/Li layer (by 0.23$-$1.54 eV), except Mn$_{\rm Li}^{3+}$ as it is more stable in the Mn/Li layer (the total-energy difference is 0.14 eV).      

{\bf Defect complexes.} In addition to the above defects, we explicitly investigate other defect complexes including but not limited to lithium divacancies (hereafter denoted as $DV_{\rm Li}$), antisite defect pairs (Mn$_{\rm{Li}}$-Li$_{\rm{Mn}}$), and a complex of Mn$_{\rm{Li}}^{+}$ and $V_{\rm{Li}}^-$ (Mn$_{\rm{Li}}^{+}$-$V_{\rm{Li}}^-$). $DV_{\rm Li}^{2-}$ is created by removing two Li$^{+}$ ions which are nearest neighbors to each other. This defect has a formation energy of 1.15$-$2.87 eV (1.54$-$3.51 eV) and a binding energy of $-$0.18 eV ($-$0.13 eV), with respect to the two isolated $V_{\rm{Li}}^-$, in \ce{LiMnO2} (\ce{Li2MnO3}). The negative binding energy indicates that the divacancies are not stable toward their isolated constituents at low lithium vacancy concentrations. The antisite pair is created by switching the positions of a Li atom and its neighboring Mn atom. In \ce{LiMnO2}, Mn$_{\rm{Li}}$-Li$_{\rm{Mn}}$ is a complex of Mn$_{\rm{Li}}^{+}$, Li$_{\rm{Mn}}^{2-}$, and $\eta^{+}$. This defect complex has a formation energy of 1.03 eV and a binding energy of 1.15 eV. In \ce{Li2MnO3}, Mn$_{\rm{Li}}$-Li$_{\rm{Mn}}$ is a complex of Mn$_{\rm{Li}}^{3+}$ and Li$_{\rm{Mn}}^{3-}$ which has a formation energy of 1.96 eV and a binding energy of 4.56 eV. The Mn$_{\rm{Li}}^{+}$-$V_{\rm{Li}}^-$ complex has a formation energy of 0.84$-$1.86 eV (1.01$-$3.84 eV) and a binding energy of 0.32 eV (0.40 eV) in \ce{LiMnO2} (\ce{Li2MnO3}); {\it cf.}~Table~\ref{tab:formenergy}. Finally, in \ce{Li2MnO3}, $V_{\rm Li}$-$V_{\rm O}$, a complex of $V_{\rm Li}^{-}$, $V_{\rm O}^{2+}$, and $\eta^{-}$, is found to have a formation energy of 2.11$-$3.06 eV, depending the chemical potentials; 2$V_{\rm Li}^{-}$-$V_{\rm O}^{2+}$, which can be interpreted as the removal of a \ce{Li2O} unit from the bulk, has a formation energy of 3.10$-$4.39 eV; and $V_{\rm Mn}$-$V_{\rm O}$, a complex of $V_{\rm Mn}^{4-}$, $V_{\rm O}^{2+}$, and two $\eta_{\rm O}^{+}$, which can be interpreted as the removal of a MnO unit from the bulk, has a formation energy of 4.21$-$6.10 eV.

For comparison, Koyama {\it et al}.~\cite{Koyama2012} in their GGA$+U$ calculations with $U=5$ eV for the Mn 3d states also find that the hole and electrons are localized in \ce{LiMnO2} and the hole is delocalized in \ce{Li2MnO3}. Their results, assuming equilibrium with \ce{O2} gas at 627$^\circ$C and 0.2 atm and \ce{Li2O}, appear to indicate that $\eta^{+}$, Li$_{\rm Mn}^{0}$, and Li$_{\rm Mn}^{-}$ have the lowest formation energies in \ce{LiMnO2}; {\it cf.}~Fig.~S2(c) in the Electronic Supplementary Information (ESI) of ref.~\cite{Koyama2012}. However, we find that their choice of the atomic chemical potentials is not suitable for \ce{LiMnO2} as it corresponds to a point on the \ce{Li2O} line in Fig.~\ref{fig;chempot}(a) that is much lower than point C and well beyond the stability region of the host compound. For \ce{Li2MnO3}, Koyama {\it et al}.'s chosen set of the chemical potentials corresponds approximately to point C in Fig.~\ref{fig;chempot}(b). Their results, {\it cf.}~Fig.~S2(d) in the ESI of ref.~\cite{Koyama2012}, appear to suggest that Li$_{i}^{+}$ and $\eta^{-}$ are the dominant defects in the material and have a formation energy of about 0.7 eV, which is in qualitative agreement with our results for these defects under the conditions at point C in Fig.~\ref{fig;chempot}(b); {\it cf.}~Table~\ref{tab:formenergy}. Using GGA calculations, Park \cite{Park2014} found that in \ce{Li2MnO3} lithium antisites Li$_{\rm Mn}$ are the dominant intrinsic defect under O-rich and Mn-poor conditions. Our results, however, show these defects always have very high formation energies, even under similar conditions such as those at points A and B in Fig.~\ref{fig;chempot}(b). We note that in these previous works, corrections for finite-cell-size effects were not included, except the ``potential alignment'' term \cite{Koyama2012,Park2014}.

Most notably absent from the previous bulk or defect calculations \cite{Koyama2009,Xiao2012,Lee2014,Park2014,Koyama2012} using GGA or GGA$+U$ is any mention of, or evidence for, the formation of bound oxygen hole polarons $\eta_{\rm O}^{+}$ in \ce{Li$_{2-x}$MnO3}. For instance, Xiao {\it et al}.'s conclusion about the partial oxidation of O$^{2-}$ during delithiation is based only on the observation that the average Bader charge at the oxygen site changes from $-$1.17 to $-$0.88 as $x$ goes from 0 to 1; {\it cf.}~Table 1 of ref.~\cite{Xiao2012}. Even in HSE06 calculations, Lee and Persson \cite{Lee2014} do not seem to observe the localization of holes at the oxygen site in \ce{Li$_{2-x}$MnO3} for $0\leq x <1$; although there appears to be evidence of O$^{-}$ for $x \geq 1$; {\it cf.}~Table 1 of ref.~\cite{Lee2014}. The results might suggest that in these previous calculations the system is not yet at its true ground state.

\subsection{Defect migration}\label{sec;migration}

\begin{figure}
\vspace{0.2cm}
\centering
\includegraphics[height=5.9cm]{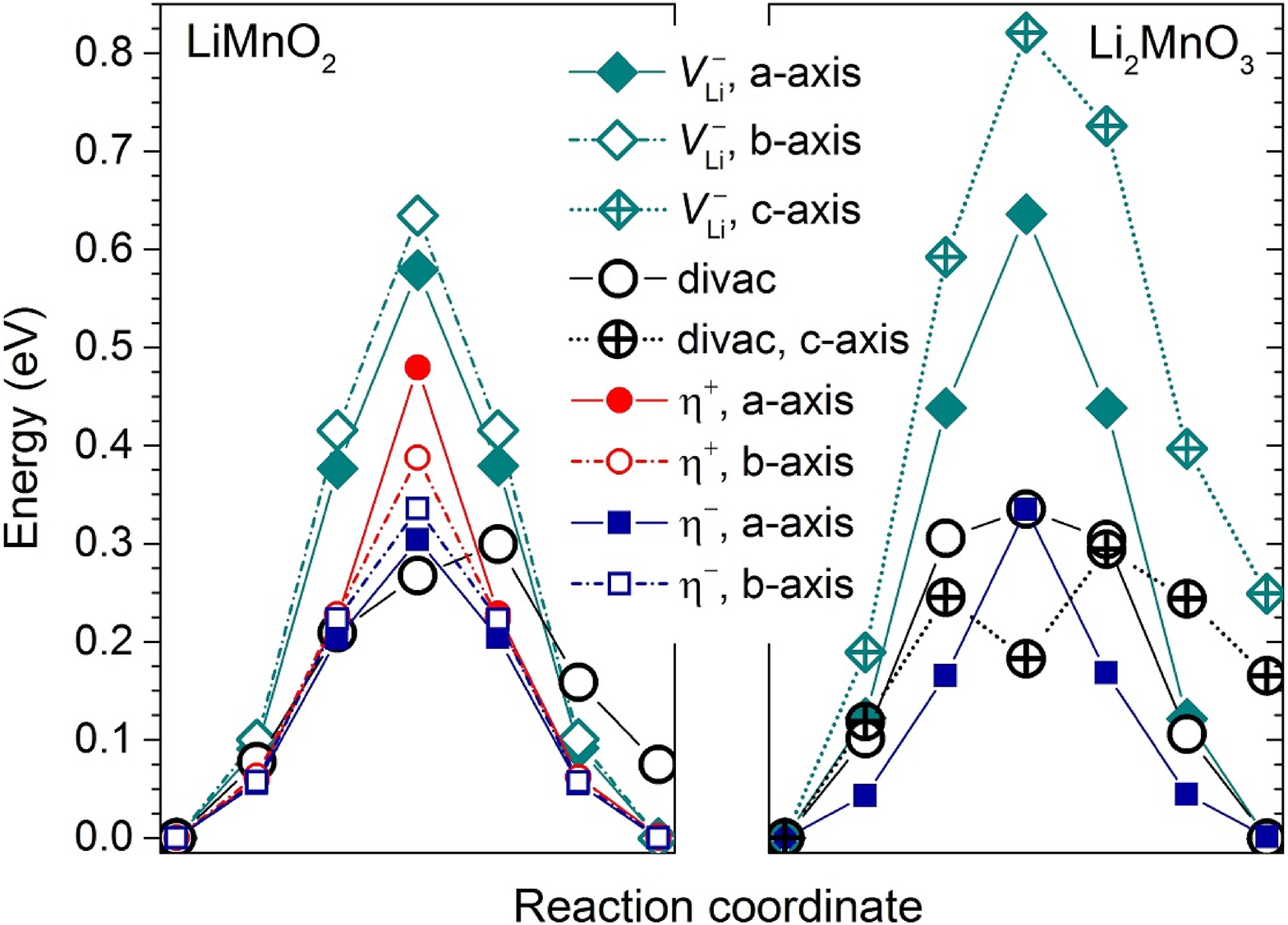}
\vspace{-0.1cm}
\caption{Calculated migration barriers of small polarons ($\eta^{+}$ and $\eta^{-}$) and lithium vacancies, via monovacancy ($V_{\rm{Li}}^-$) and divacancy (divac) mechanisms, in \ce{LiMnO2} and \ce{Li2MnO3}.}
\label{fig;migration}
\end{figure}

Figure~\ref{fig;migration} shows the migration barriers ($E_{m}$) for the small hole and electron polarons and lithium vacancies in \ce{LiMnO2} and \ce{Li2MnO3}. The migration of a polaron $\eta^{+}$ or $\eta^{-}$ between two positions $q_{\rm A}$ and $q_{\rm B}$ is described by the transfer of the lattice distortion over a one-dimensional Born-Oppenheimer surface \cite{Iordanova122,Iordanova123,Maxisch:2006p103}. We estimate the energy barrier by computing the energies of a set of cell configurations linearly interpolated between $q_{\rm A}$ and $q_{\rm B}$ and identify the energy maximum. For $V_{\rm Li}^{-}$, which migrates through a monovacancy or divacancy mechanism \cite{Hoang2014}, the barrier is estimated using the NEB method \cite{ci-neb}. It should be noted that $V_{\rm Li}^{-}$ moving in one direction is equivalent to the Li$^{+}$ ion migrating in the opposite direction. 

In \ce{LiMnO2}, we find $\eta^{+}$ has migration barriers of 0.48 and 0.39 eV along the $a$ and $b$ axes, respectively; the barriers for $\eta^{-}$ are 0.30 and 0.34 eV. For the lithium vacancies, the monovacancy mechanism gives migration barriers of 0.58 eV and 0.63 eV along the two axes, whereas the divacancy mechanism gives a lower barrier of 0.30 eV; {\it cf.}~Fig.~\ref{fig;migration}. The migration paths along the $a$ and $b$ axes are slightly different because of the Jahn-Teller distortion. In \ce{Li2MnO3}, electron polarons $\eta^{-}$ have a migration barrier of 0.33 eV in the $ab$ plane. The migration of lithium ions in the Li layer, i.e., along $a$ or $b$ axis, through monovacancy and divacancy mechanisms has energy barriers of 0.64 and 0.34 eV, respectively. Lithium ions in \ce{Li2MnO3} can also migrate across the Li and Mn/Li layers, i.e., approximately along the $c$-axis. Along this direction, the monovacancy and divacancy mechanisms give migration barriers of 0.82 and 0.29 eV, respectively; {\it cf.}~Fig.~\ref{fig;migration}. Finally, we also investigate the migration of oxygen vacancies in \ce{Li2MnO3} and find that they have very high energy barriers. For example, the migration barrier of $V_{\rm O}^{2+}$ is 1.7 eV or higher. These results are expected as the migration of oxygen vacancies involves breaking and forming Mn$-$O bonds.

For comparison, Xiao {\it et al.}~\cite{Xiao2012} in GGA$+U$ calculations with $U=5$ eV for the Mn 3d states find energy barriers of 0.61$-$0.84 eV for lithium migration in the Li layers of \ce{Li2MnO3}, which are in agreement with our value of 0.64 eV associated with the monovacancy mechanism. For the migration along the $c$ axis, they report barriers of 0.73$-$0.80 eV, also in agreement with our value of 0.82 eV reported earlier. Gao {\it et al.}~\cite{Gao2014} find comparable values in GGA$+U$ calculations with $U=4.9$ eV: 0.55$-$0.77 eV for lithium migration in the Li layer and 0.68$-$0.72 eV for migration along the $c$-axis. It should be noted that, in these works the migrating species could be $V_{\rm Li}^{0}$, instead of $V_{\rm Li}^{-}$ like in our work, and lithium migration through a divacancy mechanism is not considered.

\section{Discussion}

\subsection{Intrinsic-defect landscapes}\label{sec;landscape} 

It emerges from our results presented in Sec.~\ref{sec;formation} that certain intrinsic point defects in layered \ce{LiMnO2} and \ce{Li2MnO3} have low calculated formation energies and thus can occur with high concentrations, e.g., during materials preparation. The formation energies of some of the defects can, however, be sensitive to the chemical potentials which represent the experimental conditions.   

In \ce{LiMnO2}, the dominant point defects are $\eta^{+}$ and $V_{\rm Li}^{-}$ [under conditions at point A in the chemical-potential diagram, {\it cf.}~Fig.~\ref{fig;chempot}(a)], $\eta^{+}$ and Li$_{\rm Mn}^{2-}$ (points B and C), Mn$_{\rm Li}^{+}$ and Li$_{\rm Mn}^{2-}$ (points D$-$G), or Mn$_{\rm Li}^{+}$ and $V_{\rm Li}^{-}$ (point H); {\it cf.}~Table~\ref{tab:formenergy}. The antisite defects thus have the lowest calculated formation energies under most of the conditions. The manganese and lithium antisites can be created together in form of (i) the Mn$_{\rm Li}$$-$Li$_{\rm Mn}$ complex under conditions at points B$-$F, or (ii) the Mn$_{\rm Li}^{+}$$-$$V_{\rm Li}^{-}$ complex under conditions at points A, G, and H. In scenario (i), there are Mn$^{2+}$ ions associated with Mn$_{\rm Li}^{+}$ and Mn$^{4+}$ ions associated with $\eta^{+}$, whereas in (ii) Mn$^{2+}$ ions associated with Mn$_{\rm Li}^{+}$ are present, in addition to the Mn$^{3+}$ ions of the bulk compound. The low formation energy of Mn$_{\rm Li}^{+}$ (Li$_{\rm Mn}^{2-}$) can be partially ascribed to the small radius difference between Mn$^{2+}$ (Mn$^{3+}$) and Li$^{+}$. For reference, the Shannon ionic radii are 0.83, 0.65, and 0.76 {\AA} for high-spin Mn$^{2+}$, high-spin Mn$^{3+}$, and Li$^{+}$, respectively \cite{Shannon1976}. Our results are thus consistent with experiments showing significant cation mixing in \ce{LiMnO2}. Armstrong and Bruce \cite{Armstrong1996}, for example, report 10\% Li/Mn site disorder in \ce{LiMnO2} samples. These antisite defects are likely to act as nucleation sites for the formation of orthorhombic \ce{LiMnO2} during synthesis or spinel \ce{LiMn2O4} during electrochemical cycling, which leads to inferior cycling stability \cite{Vitins1997}.

Other intrinsic point defects in \ce{LiMnO2} include unbound, i.e., self-trapped, hole and electron polarons whose formation energies can be as low as 0.48 ($\eta^{+}$) and 0.55 eV ($\eta^{-}$); {\it cf.}~Table~\ref{tab:formenergy}. Lithium interstitials Li$_{i}^{+}$ can also occur, especially under Li-rich conditions such as at points C$-$F in the chemical-potential diagram, {\it cf.}~Fig.~\ref{fig;chempot}(a). These defects can be created in the form of Li$_{i}^{0}$, a complex of Li$_{i}^{+}$ and $\eta^{-}$. Finally, we find that oxygen and manganese vacancies are not likely to occur in the interior of \ce{LiMnO2} as their formation energies are high. These defects, however, could be energetically more favorable at the surface or interface where the lattice environment is less constrained. 

In \ce{Li2MnO3}, the lithium antisites Li$_{\rm Mn}$ have a very high formation energy ($>$2 eV), indicating that these defects are not likely to form. This high energy can be partially ascribed to the large difference in the Shannon ionic radii between Li$^{+}$ (0.76 {\AA}) and high-spin Mn$^{4+}$ (0.53 {\AA}) \cite{Shannon1976}. The formation energy of the manganese antisites Mn$_{\rm Li}$, on the other hand, can be as low as 0.71 eV under conditions at points E and F in the chemical-potential diagram; {\it cf.}~Fig.~\ref{fig;chempot}(b). It should be noted again that in Mn$_{\rm Li}^{+}$, the Mn ion is stable as high-spin Mn$^{2+}$. The low formation energy of this defect can thus be ascribed to the small ionic radius difference between Mn$^{2+}$ and Li$^{+}$, which is similar to Mn$_{\rm Li}^{+}$ in \ce{LiMnO2}. We, however, also observe that this energy is very sensitive to the atomic chemical potentials; e.g., the calculated formation energy of Mn$_{\rm Li}^{+}$ can be as high as about 3.0 eV under conditions at points A and B which represent Li-rich and Mn-poor environments. These results thus open the door to manipulating defect concentrations via defect-controlled synthesis where the experimental conditions can be tuned to reduce or enhance certain intrinsic point defects. Mn$_{\rm Li}^{+}$ can be created in form of Mn$_{\rm Li}^{0}$, a neutral complex of Mn$_{\rm Li}^{+}$ and $\eta^{-}$, especially under conditions at points D and E, or together with $V_{\rm Li}^{-}$ in form of the Mn$_{\rm Li}^{+}$$-$$V_{\rm Li}^{-}$ complex, especially under conditions at points F and G; {\it cf.}~Table~\ref{tab:formenergy}. 

Regarding other defects in \ce{Li2MnO3}, electron polarons $\eta^{-}$ have the lowest formation energy under conditions at point D in the chemical-potential diagram; {\it cf.}~Fig.~\ref{fig;chempot}(b). We note that point D represents the most reducing environment, i.e., corresponding to the lowest $\mu_{\rm O}$ value, in \ce{Li2MnO3}. Lithium interstitials Li$_{i}^{+}$ also have the lowest formation energy at point D and are the lowest-energy defects under conditions at points A$-$C. Manganese vacancies are unlikely to occur in the interior of \ce{Li2MnO3} due to their high formation energies. Oxygen vacancies may form only under highly reducing environments such as at point D where $V_{\rm O}^{+}$ has a relatively low formation energy of 1.17 eV [compared to 1.83 eV or higher at other special points in Fig.~\ref{fig;chempot}(b)]. We note that at point D the formation energies of manganese antisites are also low; {\it cf.}~Table~\ref{tab:formenergy}. Again, manganese and oxygen vacancies and any other defects can be energetically more favorable at the surface or interface. Our results for oxygen vacancies are consistent with the oxygen deficiency at the 8$j$ sites observed by Kubota {\it et al.}~\cite{Kubota2012} in \ce{Li2MnO3} when the material is synthesized in the presence of strong reducing agents. Kubota {\it et al.}~also report the presence of manganese antisites in the oxygen-deficient Li$_{2}$MnO$_{3-x}$ samples. It should be noted that there are Mn$^{3+}$ ions (in the form of $\eta^{-}$) and localized electrons $\eta_{\ast}^{-}$ associated with the oxygen vacancies, as mentioned in Sec.~\ref{sec;formation}. These species can be oxidized during the subsequent delithiation process.

Our results thus indicate that the lithium and manganese antisites in \ce{LiMnO2} cannot be eliminated just by tuning the experimental conditions, e.g., during materials preparation. An elimination of these defects may require significant changes to the chemical environment and, hence, defect-energy landscape, for example, through partial ion substitution. The concentration of manganese antisites in \ce{Li2MnO3}, on the other hand, can be significantly reduced or eliminated by preparing the material under the conditions at points A$-$C or in their nearby regions in the chemical-potential diagram; {\it cf.}~Fig.~\ref{fig;chempot}(b). It should be noted again that the conditions under which \ce{Li2MnO3} may often be prepared correspond to those approximately within the region enclosed by points C, D, and E in Fig.~\ref{fig;chempot}(b).

\subsection{Delithiation and lithiation mechanisms}\label{sec;deli}

In lithium-ion battery-electrode materials, the structure of the lithium vacancy $V_{\rm Li}^0$ and lithium interstitial Li$_{i}^{0}$ provides direct information about the intrinsic mechanisms for delithiation and lithiation, respectively \cite{Hoang2011,Hoang2014,HoangLiMn2O4}. We note that, as we are working with \ce{LiMnO2} and \ce{Li2MnO3} in their stoichiometric forms, the lithiation here should be understood as the insertion of additional lithium into the lithiated host compounds. The structure of $V_{\rm Li}^0$ in \ce{LiMnO2} indicates that for each Li atom removed from \ce{LiMnO2} electrodes during delithiation, the material is left with one negatively charged lithium vacancy $V_{\rm Li}^-$ and one hole polaron $\eta^{+}$; i.e., the extraction of lithium is associated with the oxidation of Mn$^{3+}$ to Mn$^{4+}$. The partially delithiated composition can be written as Li$_{1-x}$MnO$_{2}$ or, explicitly, Li$_{1-x}$[Mn$_{1-x}^{3+}$Mn$_{x}^{4+}$]O$_{2}$; here, for simplicity, we ignore intrinsic point defects such as lithium and manganese antisites. Regarding the lithiation process, the structure of Li$_{i}^{0}$ indicates that, for each additional Li atom inserted into \ce{LiMnO2} electrodes, the material receives one positively charged lithium interstitial Li$_{i}^+$ and one electron polaron $\eta^{-}$; i.e., the insertion of lithium is associated with the reduction of Mn$^{3+}$ to Mn$^{2+}$. The partially lithiated composition is thus Li$_{1+x}$MnO$_{2}$ or, explicitly, Li$_{1+x}$[Mn$_{1-x}^{3+}$Mn$_{x}^{2+}$]O$_{2}$. Since there are no bandlike carriers, as discussed later in Sec.~\ref{sec;conduc}, $\eta^{+}$ and $\eta^{-}$ are the electronic charge carriers in the delithiation and lithiation. These processes are thus similar to those in other electrode materials such as olivine \ce{LiFePO4} \cite{Hoang2011}, layered \ce{LiCoO2} and \ce{LiNiO2} \cite{Hoang2014}, and spinel \ce{LiMn2O4} \cite{HoangLiMn2O4}.

The deintercalation voltage [{\it cf.}~Eq.~(\ref{eq:vol1})] associated with the extraction of the first lithium from the stoichiometric \ce{LiMnO2} supercell, i.e., the creation of $V_{\rm Li}^0$, is found to be 3.62 V. This value is almost identical to the average voltage of 3.57 V computed between \ce{LiMnO2} and \ce{MnO2} in calculations using primitive cells each containing one formula unit and assuming a topotactic transition between the end compounds. Experimentally, Armstrong and Bruce report that \ce{LiMnO2} electrodes cycle between voltage limits of 3.3 to 4.3 V {\it vs.}~Li/Li$^{+}$ \cite{Armstrong1996}.

\begin{figure}
\centering
\vspace{0.2cm}
\includegraphics[height=6.8cm]{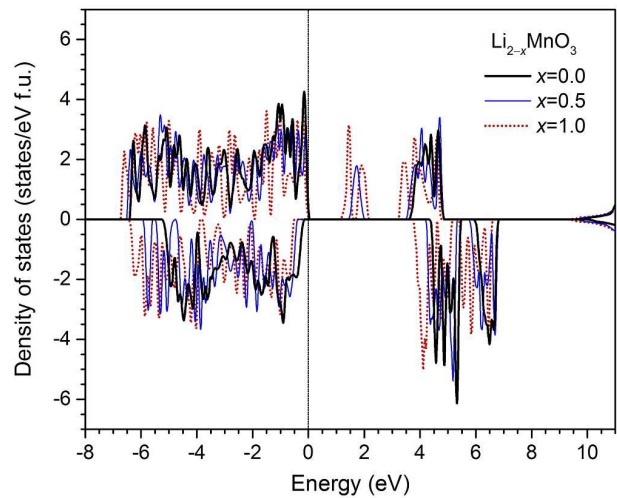}
\vspace{-0.1cm}
\caption{The electronic density of states of Li$_{2-x}$MnO$_{3}$, $x=0$, 0.5, and 1.0. The zero of energy is set to the highest occupied state.}
\label{fig;delithiated}
\end{figure}

We find that the mechanism for delithiation is completely different in \ce{Li2MnO3} where, from the structure of $V_{\rm Li}^0$, the charge-compensating defect of $V_{\rm Li}^-$ is not a hole polaron at the transition-metal site but a hole polaron at the oxygen site. For each Li atom removed from \ce{Li2MnO3} electrodes during delithiation, the material is left with one negatively charged lithium vacancy $V_{\rm Li}^-$ and one bound oxygen hole polaron $\eta_{\rm O}^{+}$; i.e., the extraction of lithium is associated with the oxidation of O$^{2-}$ to O$^{-}$. Lithium is expected to be removed first from the Li layers, at least near the start of delithiation, as the formation energy of lithium vacancies in these layers is lower than that in the Mn/Li layers; {\it cf.}~Sec.~\ref{sec;formation}. The partially delithiated composition can be written as \ce{Li$_{2-x}$MnO3} or, explicitly, Li$_{2-x}$Mn[O$_{3-x}^{2-}$O$_{x}^{-}$]. Figure \ref{fig;delithiated} shows the electronic density of states of Li$_{2-x}$MnO$_{3}$ ($x=0$, 0.5, and 1.0), calculated by using unit cells each containing two formula units. The highly localized, deep defect states at 1.2$-$2.2 eV in the band-gap region are associated with $\eta_{\rm O}^{+}$. Our results clearly indicate that the material stays nonmetallic upon delithiation, which is in contrast to previous GGA+$U$ results reported by Koyama {\it et al.}~\cite{Koyama2009} and Xiao {\it et al.}~\cite{Xiao2012} showing a metallic character. The nonmetallic character of Li$_{2-x}$MnO$_{3}$ is also observed at very low lithium vacancy concentrations, e.g., at $x=1/18$. In the delithiation process, $\eta_{\rm O}^{+}$ is, in principle, the electronic charge carrier. However, as also discussed in Sec.~\ref{sec;conduc}, the bound oxygen hole polaron is stable only in the presence of $V_{\rm Li}^-$ and thus $\eta_{\rm O}^{+}$ transport in the material near the start of delithiation is likely to be poor due to the low concentration of the lithium vacancies. Our results thus explains why it is difficult to electrochemically activate \ce{Li2MnO3}. Regarding the lithiation process, the structure of Li$_{i}^{0}$ indicates that, for each additional Li atom inserted into \ce{Li2MnO3} electrodes, the material receives one positively charged lithium interstitial Li$_{i}^+$ and one electron polaron $\eta^{-}$; i.e., the Li insertion is associated with the reduction of Mn$^{4+}$ to Mn$^{3+}$. The partially lithiated composition can thus be written as Li$_{2+x}$MnO$_{3}$ or, explicitly, Li$_{2+x}$[Mn$_{1-x}^{4+}$Mn$_{x}^{3+}$]O$_{3}$. 

The deintercalation voltage associated with the extraction of the first lithium from the bulk \ce{Li2MnO3} supercell according to the above-mentioned delithiation mechanism is found to be 5.30 V. This value is higher than the average voltage of 5.05 V computed between \ce{Li2MnO3} and \ce{LiMnO3}, by using unit cells each containing two formula units and assuming a topotactic transition between the two limits, or the average voltage of 4.90 V between Li$_{1.5}$MnO$_{3}$ and \ce{LiMnO3}. The difference indicates that the $\eta_{\rm O}^{+}$ component of $V_{\rm Li}^0$ is energetically more favorable at high lithium vacancy concentrations. We note that the calculated voltage also depends on whether lithium is removed from the interior of the material or from the surface. For example, our preliminary results show that the deintercalation voltage associated with the formation of $\eta_{\rm O}^{+}$ is 4.44 V at the (001) surface of \ce{Li2MnO3}, obtained in calculations using a semi-infinite bulk slab containing 108 atoms [similar to the supercell model presented in Fig.~\ref{fig;struct}(b)] with a vacuum region of 10 {\AA} in thickness and oxygen being the topmost layer. Also in these calculations, we find that $\eta_{\rm O}^{+}$, i.e., O$^{-}$, can be stable as free (unbound) hole polarons on the (001) surface. The calculated voltage is thus lower at the surface, as expected since the lattice environment at the surface is less constrained than in the bulk. Experimentally, \ce{Li2MnO3} is believed to be responsible for the voltage range from 4.4 to 5.0 V \cite{Yabuuchi2011}.

Regarding the other mechanisms proposed for the delithiation of \ce{Li2MnO3}, we find that the oxidation of Mn$^{4+}$ to Mn$^{5+}$ [mechanism (i)] is unlikely as our investigations find no evidence of the formation of Mn$^{5+}$ ions. The simultaneous removal of lithium and oxygen [mechanism (ii)] may occur at the surface or interface; however, oxygen is unlikely to be transported from the bulk of Li$_{2-x}$MnO$_{3}$ to the surface or interface to maintain the reaction, at least at low $x$ values, as the energy barrier associated with oxygen migration is too high; {\it cf.}~Sec.~\ref{sec;migration}. The oxidation of Mn$^{3+}$ to Mn$^{4+}$ [mechanism (iii)], where Mn$^{3+}$ ions are the preexisting defects in oxygen-deficient \ce{Li2MnO3}, is possible as discussed in Sec.~\ref{sec;landscape}; however, the oxygen deficiency is unlikely to be solely responsible for the large reversible capacity observed in experiments \cite{Pasero2005,Kubota2012}. Finally, we speculate that the oxidation of the electrolyte and exchange of H$^{+}$ for Li$^{+}$ [mechanism (iv), not addressed in our current work], if occurs, or mechanism (iii), if there are preexisting electron polarons $\eta^{-}$ (i.e., Mn$^{3+}$ ions) and/or localized electrons $\eta_{\ast}^{-}$, can help initiate the delithiation process as the mechanism associated with the oxidation of oxygen [mechanism (v), described above] is not likely to be dominant at the onset of delithiation due to the poor electronic conduction associated with $\eta_{\rm O}^{+}$.
  
\subsection{Electronic and ionic conduction}\label{sec;conduc}

In \ce{LiMnO2} or \ce{Li2MnO3}, each ionic defect has only one stable charge state, which is also called the elementary defect, except the manganese antisites Mn$_{\rm Li}$ where the Mn ion can be stable in 2+, 3+, or 4+ oxidation state. Removing (adding) electrons from (to) these elementary defects always results in defect complexes consisting of the same elementary defects and small hole (electron) polarons at the nearby lattice site(s). Even Mn$_{\rm Li}$ does not produce any shallow defect level. In addition, several positively and negatively charged defects have positive formation energies only near midgap ({\it cf.}~Fig.~\ref{fig;formenergy}), making them perfect charge compensators. Any attempt to deliberately shift the Fermi level of the system from $\mu_{e}^{\rm int}$ to the VBM or CBM will thus result in the charged defects having negative formation energies, i.e., the intrinsic defects will form spontaneously and counteract the effects of shifting \cite{Hoang2011,Hoang2014,HoangLiMn2O4,Hoang2012274}. Clearly, intrinsic point defects in \ce{LiMnO2} and \ce{Li2MnO3} cannot act as sources of bandlike electrons and holes, and the material cannot be doped n- or p-type. The electronic conduction therefore occurs through hopping of hole and/or electron polarons. The ionic conduction, on the other hand, proceeds via lithium monovacancy or divacancy migration. 

Charge-carrying defects in the electronic and ionic conduction can be thermally and/or athermally activated. If the activation is predominantly thermal, the effective activation energy for conduction is the sum of the defect-formation energy and migration barrier, i.e., $E_{a} = E^{f} + E_{m}$; whereas if it is predominantly athermal, the effective activation energy is dominated by the migration barrier part, i.e., $E_{a} \sim E_{m}$ \cite{Hoang2014}. In stoichiometric \ce{LiMnO2} and at high temperatures, the lower bound of the activation energy associated with $\eta^{+}$ for electronic conduction is estimated to be 0.87 eV, which is its lowest formation energy (0.48 eV) plus the lowest migration barrier value (0.39 eV), and that associated with $\eta^{-}$ is $0.55 + 0.30 = 0.85$ eV; {\it cf.}~Table~\ref{tab:formenergy}. For the ionic conduction, the activation energy associated with $V_{\rm Li}^{-}$ is as low as $0.48 + 0.58 = 1.06$ eV. Here, the concentration of thermally activated lithium vacancies is likely to be quite low and lithium migration may proceed via the monovacancy mechanism. In partially delithiated Li$_{1-x}$[Mn$_{1-x}^{3+}$Mn$_{x}^{4+}$]O$_{2}$, where preexisting (athermal) $\eta^{+}$ and $V_{\rm Li}^{-}$ are the predominant charge-carrying defects, the activation energy for electronic conduction can be as low as 0.39 eV, i.e., the lowest migration barrier of $\eta^{+}$; and that for ionic conduction is 0.30 eV, i.e., the lithium divacancy migration barrier. In this case, the concentration of lithium vacancies is likely to be so high that the divacancy mechanism is more favorable.

Similarly, the lower bound of the activation energy for electronic conduction associated with $\eta^{-}$ in stoichiometric \ce{Li2MnO3} and at high temperatures is estimated to be $0.75 + 0.33 = 1.08$ eV; {\it cf.}~Table~\ref{tab:formenergy}. For the ionic conduction, the lower bound of the activation energy is estimated to be 1.32 eV, the sum of the lowest formation energy (0.71 eV) and migration barrier (0.61 eV) of $V_{\rm Li}^{-}$. Here, again, we assume that the concentration of lithium vacancies is low and thus the monovacancy migration mechanism is predominant. In partially delithiated Li$_{2-x}$Mn[O$_{3-x}^{2-}$O$_{x}^{-}$], athermal $\eta_{\rm O}^{+}$ and $V_{\rm Li}^{-}$ are the predominant charge-carrying defects. The lower bound of the activation energy for ionic conduction is estimated to be 0.29 eV, which is the lowest migration barrier of $V_{\rm Li}^{-}$ associated with the divacancy mechanism. The oxygen hole polaron $\eta_{\rm O}^{+}$ that is bound to $V_{\rm Li}^{-}$ can, in principle, contribute to the electronic conduction. However, a high concentration of $V_{\rm Li}^{-}$ would be needed in order to form the percolation pathways for $\eta_{\rm O}^{+}$ diffusion. The difficulty in $\eta_{\rm O}^{+}$ transport, especially at low lithium vacancy concentrations, is likely to result in poor electronic conduction and electrochemical performance.

Massarotti {\it et al.}~\cite{Massarotti199794} report that \ce{Li2MnO3} is an insulator with a very small electrical conductivity ($< 10^{-10}$ S/cm). This result is consistent with the high electronic and ionic activation energies associated with $\eta^{-}$ and $V_{\rm Li}^{-}$, respectively, estimated from our calculations for the stoichiometric compound. It may also indicate that in their measurements $\eta_{\rm O}^{+}$ does not effectively contribute to the total conductivity. Nakamura {\it et al.}~\cite{Nakamura2010}, on the other hand, report an activation energy of 0.46 eV for Li$^{+}$ hopping in \ce{Li2MnO3} samples prepared by ball milling and sintered at 900$^\circ$C. This value is higher than the calculated migration barrier of $V_{\rm Li}^{-}$ associated with the lithium divacancy mechanism but lower than that associated with the monovacancy mechanism, suggesting that there is a high concentration of athermal lithium vacancies in their samples. It should be noted that a comparison between calculated and measured values is usually complicated by the fact that the latter can be sensitive to the synthesis conditions and the measurements.

We anticipate that the electronic conduction associated with $\eta_{\rm O}^{+}$ can be more effective at the surface or interface. Diffusion length shortening can also help improve the $\eta_{\rm O}^{+}$ transport. In fact, electrochemical performance improvement through nanostructuring has been explored and shows positive results which may be ascribed, at least partially, to an improvement in the electronic conduction. \ce{Li2MnO3} nanoparticles and nanowires, in particular, are reported to exhibit superior electrochemical properties compared to their bulk counterpart \cite{Jain2005,Yu2009,Lim2012,Wu2014}. Finally, the electronic conduction can also be improved via ion substitution. A partial substitution of Mn$^{4+}$ in \ce{Li2MnO3} with electrochemically active metal ions would introduce an additional electronic conduction and charge-compensation mechanism that is highly needed at the start of the delithiation process. Indeed, this appears to be the case in \ce{Li2Ru$_{1-y}$Mn$_y$O3} ($0.2 \leq y \leq 0.8$) cathode materials where Ru$^{4+}$ ions can be oxidized to Ru$^{5+}$ \cite{Sathiya2013}.

\section{Conclusions}

We carry out a comprehensive study of the bulk properties and defect physics in layered lithium manganese oxide cathode materials using a hybrid DFT/Hartree-Fock method. We find that layered \ce{LiMnO2} has a Jahn-Teller distorted, monoclinic structure and its energy at 0 K is degenerate with that of orthorhombic \ce{LiMnO2}. An analysis of the electronic structure in \ce{LiMnO2} shows that the contribution from Mn 3d states to the VBM is larger than that from O 2p states in each oxygen atom; whereas the VBM in \ce{Li2MnO3} is predominantly O 2p states. This difference between the two compounds results in different defect physics, particularly in those intrinsic point defects whose formation involves removing electrons from the top of the valence band.

Manganese antisites are found to have low formation energies and thus can occur with high concentrations in layered \ce{LiMnO2}; the low energies can be partially ascribed to the small ionic radius difference between high-spin Mn$^{2+}$ and Li$^{+}$. These antisites can act as nucleation sites for the formation of orthorhombic or spinel phases during synthesis or electrochemical cycling. An elimination of these antisite defects would require significant changes to the chemical environment such as through ion substitution. \ce{Li2MnO3} can also have high concentrations of manganese antisites; however, unlike in \ce{LiMnO2}, they can be eliminated by tuning the experimental conditions. Other intrinsic point defects may also occur and have an impact on the materials' properties and functioning. 

A detailed analysis of the formation of lithium vacancies in layered \ce{LiMnO2} indicates that the delithiation process is associated with the oxidation of Mn$^{3+}$ to Mn$^{4+}$, leading to the formation of small hole polarons $\eta^{+}$ at the transition-metal site. In Li-excess \ce{Li2MnO3}, the intrinsic mechanism for lithium extraction is found to be associated with the oxidation of O$^{2-}$ to O$^{-}$, leading to the formation of bound hole polarons $\eta_{\rm O}^{+}$ at the oxygen site. Other delithiation mechanisms can also occur in \ce{Li2MnO3} and may be dominant near the start of the delithiation process; however, it is this intrinsic mechanism that can explain the large reversible capacity observed in experiments. We also find that in both compounds the intrinsic point defects cannot act as sources of bandlike electrons and holes and the materials cannot be doped n- or p-type. The electronic conduction proceeds through hopping of hole and/or electron polarons; the ionic conduction occurs through lithium monovacancy and/or divacancy migration mechanisms. 

Since hole polarons $\eta_{\rm O}^{+}$ are not stable in the interior of \ce{Li2MnO3} in the absence of negatively charged lithium vacancies, the electronic conduction at low lithium vacancy concentrations is likely to be poor due to the lack of percolation pathways for $\eta_{\rm O}^{+}$ diffusion. We suggest that one can improve the electronic conduction and, hence, the electrochemical performance of \ce{Li2MnO3} through nanostructuring and/or ion substitution. Finally, the results and discussion presented in this work can also shed light on the electrochemical properties of \ce{Li2MnO3}-based or related materials, opening the door to utilizing the oxygen oxidation mechanism for high-capacity battery electrodes.

\begin{acknowledgments}

We thank Michelle Johannes for useful discussions and Steve Erwin for critical reading of the manuscript. This work was supported by the U.S.~Department of Energy (Grant No.~DE-SC0001717) and the Center for Computationally Assisted Science and Technology (CCAST) at North Dakota State University.

\end{acknowledgments}


%
\end{document}